\documentclass[11pt]{article}
\usepackage[centertags]{amsmath}
\usepackage[latin1]{inputenc}
\usepackage{amssymb,amsthm}
\usepackage{amsmath}
\usepackage{amsfonts}
\usepackage{textcomp}
\usepackage{newlfont}
\usepackage{graphicx}
\usepackage[dvips]{color}

\newtheorem{proposition}{Proposition}[section]

\newtheorem{rem}{Remark}[section]
\newtheorem{ass}{Assumption}[section]
\newcommand{\e}{\mathrm{e}}
\newcommand{\E}{\mathbb{E}}

\newcommand{\R}{\mathbb{R}}

\newcommand{\ci}{\mathrm{i}}
\newcommand{\eps}{\epsilon}

\def\P{{\mathcal{P}}}
\def\Q{{\mathcal{Q}}}
\def\half{\frac{1}{2}}
\def\Var{\mathrm{VaR}}
\def\AVar{\mathrm{AVaR}}

\title{Computing Quantiles in Regime-Switching Jump-Diffusions with Application to Optimal Risk Management:
a Fourier Transform Approach}
\author{Alessandro Ramponi\footnote{The financial support of the Research Grant
\emph{PRIN 2008, Probability and Finance, Prot. 2008YYYBE4}, is
gratefully acknowledged. } \\ \\
Department of Economics and Finance \\
University of Roma - Tor Vergata\\
via Columbia, 2 - 00133 Roma, Italy \\
e-mail: ramponi@economia.uniroma2.it}

\date{}

\begin{document}

\maketitle

\begin{abstract}
In this paper we consider the problem of calculating the quantiles
of a risky position, the dynamic of which is described as a
continuous time regime-switching jump-diffusion, by using Fourier
Transform methods. Furthermore, we study a classical option-based
portfolio strategy which minimizes the Value-at-Risk of the hedged
position and show the impact of jumps and switching regimes on the
optimal strategy in a numerical example. However, the analysis of
this hedging strategy, as well as the computational technique for
its implementation, is fairly general, i.e. it can be applied to
any dynamical model for which Fourier transform methods are
viable.

\medskip


%\medskip
%\noindent \textbf{JEL Classification}: E43, G12, G13


%\medskip
%\noindent \textbf{Mathematics Subject Classification (1991)}:
%90A09, 60G35
\end{abstract}

\section{Introduction}
Quantiles computation of profit and losses of a given financial
position is a basic task for measuring and managing portfolio
market risks. The dynamic of the driving risk factors, as well as
the chosen risk measure are the basic ingredient for the analysis
of any risk management strategy. In this paper the model we
consider for the risky position is of the form $S(t) = s_0
\e^{X(t)}$, where $X(t)$ is specified on a filtered probability
space $(\Omega, \{\mathcal{F}_t \}, \mathcal{F}, \mathcal{P})$ as
a jump-diffusion whose parameters change over time, driven by a
continuous time and stationary Markov Chain on the finite state
space $\mathcal{S} = \{1,2,\ldots,M \}$, representing the
unobserved \emph{state of the world}. In fact, empirical studies
on the behavior of financial markets show the ability of
regime-switching models to capture some peculiarities in the
observed data, as firstly highlighted in the seminal paper by
Hamilton \cite{Ha}. Since then there has been a growing effort in
applying switching models to a wide class of financial and/or
economic problems, such as time series analysis, portfolio theory,
derivative pricing and risk management. On the other hand, the
necessity of including jumps in the underlying models to provide
better representation of their dynamical properties is widely
recognized (see e.g. \cite{CT}). Regime-switching jump-diffusions
turns out to be an appealing and flexible class of dynamic models.
%able to reproduce empirical stylized facts about observed data,
%such as volatility clustering and heavy tails property of
%distributions.

Among many different risk measures proposed in the literature,
Value-at-Risk, although sharply criticized for the lack of
sub-additivity and its inability to quantify the severity of an
exposure to rare events, has been adopted as a benchmark in the
financial industry and for regulatory purposes. It plays a central
role in banking regulation and internal risk management, mainly
due to its simplicity. We therefore take the VaR as a starting
point of our analysis of risk management strategies. The
computation of VaR in Regime-Switching models has been considered
by several authors mainly in discrete-time setting (see e.g.
\cite{BP} or more recently \cite{KK}, \cite{Taa09}). Here we
consider this problem directly in the continuous time framework:
as a matter of fact, the required computations can be very
efficiently implemented with the help of Fourier Transform methods
(see e.g. \cite{Ram10}). The use of this kind of technique for the
analytical calculation of VaR has been considered in Duffie and
Pan \cite{DP} who exploited the classical characterization of the
distribution function in terms of the Fourier inversion of its
characteristic function. The use of Generalized Fourier Transform
and the FFT algorithm is more recent: see Le Courtois and Walter
\cite{LCW} who calculate the VaR for the Variance Gamma (VG) model
and Kim et al. \cite{KRBF}, Scherer et al. \cite{SRKF}, who
consider the class of tempered stable and infinitely divisible
distributions.

As an application to risk management, we investigate the influence
of jumps and switching regimes on the exposure to an underlying
risky asset. More precisely we study a classical hedging policy
based on options followed by an institutional manager whose aim is
to minimize the VaR of a position. This type of analysis has been
initiated by \cite{ABRV} a decade ago for a portfolio made by a
risky asset following a log-normal random dynamic, and hence
analytically solved in a Black-Scholes setting. More recently, it
has been considered for a bond portfolio in \cite{DEHV},
\cite{ADHV}, \cite{ARS}.

By taking the VaR as the risk measure for potential losses $L$ of
a portfolio at a given level $\alpha$ (i.e. the value
$\Var_{\alpha}(L)$ such that $\mathrm{Prob}\{ L >
\Var_{\alpha}(L)\} = \alpha$), the strategy considered consists on
minimizing the VaR of the option-hedged portfolio $L^{h,K}$ with
respect to the strike price and the quantity of the put option
written on the risky asset, subject to a budget constraint: in
other words, we hedge the risky position by buying a fraction
$h\in (0,1]$ of a put option with maturity $T$ and strike price
$K$, but what $K$ and $h$? The optimal hedging strategy is
therefore given by the solution of the following \emph{program}:
$$
\left \{
\begin{array}{l}
  \min_{K,h} \hbox{VaR}_{\alpha}(L^{h,K}) \\ \\
  h \Pi^P_0(K,T) = C, \ \ \ h \in
(0,1] \\
\end{array} \right.
$$
$\Pi^P_0(K,T)$ being the time $0$ put price with strike $K$ and
maturity $T$. To implement the strategy, it is therefore needed
\textit{i)} the calculation of the VaR for the risky asset and
\textit{ii)} the corresponding value of a put option
$\Pi^P_0(K,T)$. Under the classical no-arbitrage assumption, the
price of the put option can be represented as the expected value
of the payoff with respect to a risk-neutral probability (see e.g.
\cite{Bjork}). Conversely, VaR is obtained under the objective or
historical probability measure. Both steps can be efficiently
faced with the Fourier transform technique.

The paper is organized as follows: we firstly derive the
optimality conditions for the VaR minimizing strategy (Section 2)
and then we specify the Fourier Transform technique for
calculating quantiles and put/call option prices in a very general
setting (Section 3). In Section 4 we introduce the
regime-switching dynamic model, its generalized characteristic
function and the main change-of-measure result for switching from
the historical to the risk-neutral probability. Finally, in
Section 5 some numerical experiments are reported to show the
impact of jumps and regime-switching on the process quantiles and
on the optimal hedging strategy. We use in particular a simple
two-state model with gaussian jumps and quantify in such a case
what is the effect of a wrong model choice.

\section{VaR and optimal risk management}

To measure the risk of a financial position the quantiles of its
distribution function are commonly used. Given a confidence level
$\alpha \in (0,1)$, the set of $\epsilon$-quantiles of the random
variable $Y$ is the interval $[q^-_{\alpha}(Y), q^+_{\alpha}(Y)]$
where
$$
q^-_{\alpha}(Y) = \inf \{q \in \mathbb{R} | P(Y\leq q) \geq \alpha
\}, \ \ q^+_{\alpha}(Y) = \inf \{q \in \mathbb{R} | P(Y\leq q) >
\alpha \}.
$$

For a random variable having continuous and strictly increasing
distributions function $F_Y(y)$, $q^-_{\alpha}(Y) =
q^+_{\alpha}(Y) \equiv q_{\alpha}(Y)$ and $q_{\alpha}(Y) =
F^{-1}_Y(\alpha)$, i.e. it solves the equation
$$
P( Y \leq q_{\alpha}(Y)) = \alpha.
$$

Here we take the portfolio loss $L$ to describe a financial
position in a fixed time interval and, in order to simplify
notations, we assume in this section that $L$ has a continuous and
strictly increasing distributions function. The Value-at-Risk at
level $\alpha$ is defined as
$$
\Var_{\alpha}(L) \equiv  \inf \{q \in \mathbb{R} | P(L \leq q)
\geq  \alpha \}.
$$

Let $S_t$ be the value of the risky asset, $t \in [0,T]$ and $r$
be the risk-free rate, that without loss of generality we consider
fixed in the period: we define the loss at time $0$ of such a
position as
$$
L^u = S_0-\e^{-rT}S_T.
$$

Then
$$
\Var_{\alpha}(L^u) =  S_0 - \e^{-rT} q_{1-\alpha}(S_T).
$$

Let us now consider a classical hedging problem in which an
institution has an exposure to a risky asset $S_t$ and decide to
hedge such an exposure in the interval $[0,T]$ by buying a
fraction $h \in (0,1]$ of an European put option on the asset with
maturity $T$ and strike price $K$. Analogously to the situation
considered by Ahn et al. (1999), we take as the hedged position
the portfolio composed by the risky asset and the put option: the
loss of the hedged portfolio at time $0$ is therefore
$$
L^{h,K} = S_0+h \Pi^P_0(K,T) - \e^{-rT}(S_T+h(K-S_T)^+)
$$
where $\Pi^P_t(K,T)$ is the price of the put option at time $t$.
By defining the strictly increasing function
$$
g(u) = u - h (u-\bar K)^+ +h \Pi^P_0(K,T)
$$
where $\bar K = S_0-\e^{-rT}K$, it is immediately seen that
$$
L^{h,K} = g(L^u);
$$
therefore
$$
\Var_{\alpha}(L^{h,K}) = g(\Var_{\alpha}(L^u)) = S_0 - \e^{-rT}
q_{1-\alpha}(S_T) + h \Pi^P_0(K,T) - \e^{-rT} h
(K-q_{1-\alpha}(S_T))^+
$$
$$
= \Var_{\alpha}(L^u) + h \Pi^P_0(K,T) - \e^{-rT} h
(K-q_{1-\alpha}(S_T))^+.
$$

Let us firstly notice that if $K \leq q_{1-\alpha}(S_T)$, then
$$
\Var_{\alpha}(L^{h,K}) > \Var_{\alpha}(L^u)
$$
since $\Pi^P_0(K,T)>0$. Therefore the optimal hedging strategy is
given by the following problem:
\begin{equation} \label{constr_problem}
\left \{
\begin{array}{l}
  \min_{K,h} \Var_{\alpha}(L^u) + h \Pi^P_0(K,T) - \e^{-rT} h
(K-q_{1-\alpha}(S_T)) \\ \\
   h \Pi^P_0(K,T) = C, \\ \\
   h \in (0,1], \ \ K > q_{1-\alpha}(S_T),
\end{array} \right.
\end{equation}
$C$ being the budget constraint. Since $h = C / \Pi^P_t(K,T)$, the
optimality first order condition for $K$ is given by the following
non-linear equation:
\begin{equation} \label{kappastar_eq}
\Pi^P_0(K,T) = (K-q_{1-\alpha}(S_T) ) \frac{\partial}{\partial K}
\Pi^P_0(K,T).
\end{equation}

Assuming that (\ref{kappastar_eq}) has a solution $K^* >
q_{1-\alpha}(S_T)$ and the twice differentiability of the price
functional we can prove that this is actually a minimum since
$$
\frac{\partial^2 \Var_{\alpha}(L^{h,K})}{\partial K^2}  = \frac{
\e^{-rT} C (K^*-q_{1-\alpha}(S_T))}{\Pi^P(K^*,T)^2}
\frac{\partial^2 \Pi^P}{\partial K^2}(K^*,T) > 0
$$
by the convexity of the price functional w.r.t. the strike.
Correspondingly, the optimal amount of the hedging put option is
\begin{equation} \label{hstar_eq}
h^* = \frac{C}{\Pi^P_0(K^*,T)}.
\end{equation}

We now assume the following:

{\ass The price of the put option can be represented as the
discounted expected value of the payoff at time $T$ under a
risk-neutral measure $\Q$:
$$
\Pi^P_t(K,T) = \e^{-rT} \E^{\Q}[(K-S_T)^+].
$$
Furthermore, let $F_S(s) = \Q(S_T\leq s)$ be the cumulative
distribution function (cdf) of the random variable $S_T$ under
such a measure: hence
\begin{eqnarray*}
\Pi^P_t(K,T) & = & \e^{-rT} \int_{-\infty}^{+\infty} (K-s)^+
dF_S(s) =\e^{-rT}\left( K \int_{-\infty}^{K} dF_S(s) -
\int_{-\infty}^{K} s dF_S(s) \right) \\
 & = & \e^{-rT}\left( K \Q(S_T \leq K) -
\int_{-\infty}^{K} s dF_S(s) \right)
\end{eqnarray*}
and
$$
\frac{\partial}{\partial K} \Pi^P_t(K,T) = \e^{-rT} \Q(S_T\leq K).
$$
\label{assunz1}}

We can finally prove the following property:

{\proposition If $K^*> q_{1-\alpha}(S_T)$, then
$\Var_{\alpha}(L^{h^*,K^*}_0) < \Var_{\alpha}(L^u)$.

\proof Since $K^*$ and $h^*$ are characterized through
(\ref{kappastar_eq}) and (\ref{hstar_eq}), we get
$$
\Var_{\alpha}(L^{h^*,K^*}_0) = \Var_{\alpha}(L^u) + C - \e^{-rT}
h^* (K^*-q_{1-\alpha}(S_T))
$$
$$
= \Var_{\alpha}(L^u) + \frac{C}{\Pi^P_t(K^*,T)} \left (
\Pi^P_t(K^*,T) - \e^{-rT} (K^*-q_{1-\alpha}(S_T) ) \right )
$$
$$
= \Var_{\alpha}(L^u) + \frac{C}{\Pi^P_t(K^*,T)}
(K^*-q_{1-\alpha}(S_T) ) \left(\frac{\partial}{\partial K}
\Pi^P_t(K^*,T) - \e^{-rT} \right).
$$

From Assumption \ref{assunz1}, we have
$$
\frac{\partial}{\partial K} \Pi^P_t(K,T) - \e^{-rT}  = \e^{-rT}
(\mathbb{Q}(S_T\leq K) -1) < 0.
$$
Therefore
$$
\Var_{\alpha}(L^{h^*,K^*}_0) = \Var_{\alpha}(L^u) +
\frac{C}{\Pi^P_t(K^*,T)} (K^*-q_{1-\alpha}(S_T) ) \e^{-rT}
(\mathbb{Q}(S_T\leq K^*) -1) < \Var_{\alpha}(L^u).
$$
\endproof}

\bigskip

{\rem \label{remPQ} Notice that the optimality condition
(\ref{kappastar_eq}) under Assumption \ref{assunz1} becomes
$$
\e^{-rT}\left( K \Q(S_T \leq K) - \int_{-\infty}^{K} s dF_S(s)
\right) = (K-q_{1-\alpha}(S_T) ) \e^{-rT} \Q(S_T\leq K)
$$
which simplifies to
\begin{equation}\label{Kstareq0}
\frac{1}{\Q(S_T\leq K)}\int_{-\infty}^K s dF^Q_S(x) =
q_{1-\alpha}(S_T)
\end{equation}
and depends on both the objective and the risk neutral
distributions $\P$ and $\Q$. Furthermore, it easily seen that the
l.h.s. is equal to the conditional expectation $\E^Q[S_T | S_T
\leq K]$ which is an increasing function of $K$ bounded by
$\E^{\Q}[S_T]$. Therefore, the eq. (\ref{Kstareq0}) has a unique
solution if and only if $q_{1-\alpha}(S_T) <\E^{\Q}[S_T]$. }

\bigskip

The minimum VaR as a function of the budget $C$ is therefore
\begin{equation} \label{Var_effront}
\Var^* =\Var_{\eps}(V) + C \frac{\Pi(K^*,T) \e^{rT}-
(K^*-q_X^+(\eps))}{\Pi(K^*,T)},
\end{equation}
which is a linear function with negative slope. In the plane
budget-risk the previous equation describe an \emph{efficient
frontier} giving for each level of the budget $C$ the minimum VaR.

\bigskip

{\rem It is easy to show that the problem of looking for the
hedging strategy with the minimum cost $C (h,K)= h \Pi(K,T)$ for a
target level $v$ of VaR, results in the same first order
optimality condition for $K$ and that we get the same linear
efficient frontier
$$
C^* = h^* \Pi(K^*) = (v - \Var_{\eps}(V)) \frac{\Pi(K^*)}{\Pi(K^*)
\e^{rT} - (K^*-q_X^+(\eps))},
$$
in the budget-risk plane.
}

\section{The Fourier transform method}

Fourier transform methods are efficient techniques emerged in
recent years as one of the main methodology for the evaluation of
derivatives. In fact, the no-arbitrage price of an european style
contingent claim can be represented as the (conditional)
expectation of the derivative payoff under a proper risk-neutral
measure (see e.g. \cite{Bjork}). These methods essentially consist
on the representation of such an expectation as a "convolution" of
two Generalized Fourier Transforms. Since the value of most
derivatives depend on a trigger parameter, two main variants have
been developed depending on which variable of the payoff is
transformed into the Fourier space. Here we consider the technique
introduced in \cite{Carr99} which consider the generalized Fourier
transform with respect to the trigger parameter.

More formally, let $\Pi(S,K)$ be the payoff at maturity of the
derivative: for example, $\Pi(S,K)=(K-S)^+$ is the payoff of the
put option. The no-arbitrage price is therefore given by
$$
\Pi_0 = \e^{-r T} \E^{\Q}[\Pi(S_T,K)].
$$

Due to the exponential structure of typical underlying dynamics of
the form $S_t=s_0 \exp(X_t)$, it is convenient to represent the
payoff with respect to the new variables $X_T=\log(S_T)-\log(s_0)$
and $k=\log(K)$, in such a way $\Pi(S_T,K) = \Pi(s_0 \exp(X_T),
\exp(k)) = \Pi(X_T+\log(s_0), \log(K))$.

Therefore, let us denote with $\Pi(x,k)$ an arbitrary payoff
function and with $\hat \Pi_x(z)$ its generalized Fourier
transform (GFT) w.r.t. $k$, that is
$$
\hat \Pi_x(z) = \int_{\mathbb{R}} \e^{\ci z k} \Pi(x,k) dk, \ \ \
z \in \mathbb{C};
$$
under proper regularity conditions (see e.g. \cite{Lee04}),
Fourier inversion gives
$$
\Pi(x,k) =  \frac{1}{2\pi} \int_{\ci \nu - \infty}^{\ci \nu
+\infty} \e^{-\ci z k} \hat \Pi_x(z) dz,
$$
in some strip of $\mathbb{C}$, from which
$$
B_T \Pi_0 \equiv \E[\Pi(X(T)+\log(s_0),k)] = \int_\mathbb{R}
\Pi(x,k) \mathcal{Q}(dx)
$$
$$
= \int_\mathbb{R} \frac{1}{2\pi} \int_{\ci \nu - \infty}^{\ci \nu
+\infty} \e^{-\ci z k} \hat \Pi_x(z) dz \mathcal{Q}(dx) =
\frac{1}{2\pi} \int_{\ci \nu - \infty}^{\ci \nu +\infty} \e^{-\ci
z k}  \int_\mathbb{R} \hat \Pi_x(z) \mathcal{Q}(dx) dz
$$
\bigskip
$$
= \frac{1}{2\pi} \int_{\ci \nu - \infty}^{\ci \nu +\infty}
\e^{-\ci z k} \E^{\mathcal{Q}}[\hat \Pi_{X(T)+\log(s_0)}(z)] dz.
$$

In order to implement our program, we need to evaluate
\begin{enumerate}
\item the VaR of the hedged position: this step require to solve w.r.t. $v$
the equation $\P\{ S_T < v \} = \alpha$;
\item the value of a put option
$\Pi^P_0(k,T)= \e^{-r T} \E^Q[(\e^{k}-S_0 \e^{X_T})^+]$.
\end{enumerate}
Let us consider the following "payoff" functions,
$$
\Phi_1(x,k) = (\e^{k}-\e^{x})^+, \ \ \mbox{and} \ \ \Phi_2(x,k) =
\mathbb{I}_{\{x \leq k\}},
$$
in such a way $\Pi^P_0(\kappa,T)= \e^{-r T}
\E^Q[\Phi_1(X_T+\log(S_0),k)]$ and $\P\{ S_T < v \} = \P \{ X_T <
k \} = \E[\Phi_2(X_T,k)]$, with $k= \log(v/S_0)$. Their GFT w.r.t.
the trigger parameter $k$ are
$$
\hat{\Phi}_1(x,z) = \frac{\e^{x(1+\ci z)}}{\ci z - z^2}, \ \ \ \nu
> 1 \ \ \mbox{and} \ \ \hat{\Phi}_2(x,z) = \frac{\ci}{z} \e^{\ci x z}, \ \ \
\nu > 0
$$
giving therefore the formulas
\begin{eqnarray}\label{put_formula}
\Pi^P_0(k,T) & = & \e^{-r T} \frac{1}{2 \pi} \int_{\ci \nu -
\infty}^{\ci \nu + \infty} \e^{- \ci k z} \frac{\e^{\ci
\log(S_0)(z-\ci)} \phi_X(z-\ci)}{\ci z - z^2} dz,  \ \ \ \nu
> 1 \nonumber \\
& = & \e^{-r T} \frac{\e^{\nu k} S_0^{1-\nu}}{\pi} \Re
\left(\int_{0}^{+ \infty} \e^{-\ci u (k-\log(S_0))}
\frac{\phi_X(u+\ci(\nu-1))}{\nu^2-u^2-\nu + \ci u (1-2 \nu)} du
\right),
\end{eqnarray}
and
\begin{eqnarray}\label{proba_formula}
\P\{ S_T < v \} & = & \frac{\ci}{2 \pi} \int_{\ci \nu -
\infty}^{\ci \nu + \infty} \e^{- \ci z \log(v/S_0)} \frac{\phi_X(z)}{z} dz, \ \ \ \nu > 0 \nonumber \\
& = & \frac{(v/S_0)^{\nu}}{\pi} \Re \left( \int_{0}^{+\infty}
\e^{- \ci u \log(v/S_0)} \frac{\phi_X(u + \ci \nu)}{\nu-\ci u} du
\right)
\end{eqnarray}
$\phi_X(z)$ being the GFT of the process $X_T$ under the
appropriate measure. If this is a regular functions in a properly
defined strip of $\mathbb{C}$, the transform method can be applied
in both cases (see Lee (2004)).

\bigskip

Since under the Assumption (\ref{assunz1}) the optimality
condition is
$$
(K-q_{1-\alpha}(S_T) ) = \frac{\Pi^P_0(K,T)}{
\frac{\partial}{\partial K} \Pi^P_0(K,T)} \equiv
\frac{\Pi^P_0(K,T)}{\e^{-rT} \Q(S_T\leq K)}.
$$
the optimal hedging strategy is then implemented by running

\begin{enumerate}
\item root search algorithm to find the value $q^*$ solution of
$$
\frac{(q^*/S_0)^{\nu}}{\pi} \Re \left( \int_{0}^{+\infty} \e^{-
\ci u \log(q^*/S_0)} \frac{\phi_X^{\mathcal{P}}(u + \ci
\nu)}{\nu-\ci u} du \right) = \alpha, \ \ \ \nu > 1;
$$
\item root search algorithm to find the value $K^*$ solution of
$$
(K^* - q^*)= \frac{S_0 \Re \left(\int_{0}^{+ \infty} \e^{-\ci u
(\log(K^*/S_0))}
\frac{\phi_X^{\mathcal{Q}}(u+\ci(\nu-1))}{\nu^2-u^2-\nu + \ci u
(1-2 \nu)} du \right)}{\Re \left(\int_{0}^{+ \infty} \!\! \e^{-\ci
u (\log(K^*/S_0))} \frac{\phi_X^{\mathcal{Q}}(u+\ci(\bar
\nu-1))}{\bar \nu-\ci u} du \right)}, \ \ \nu>1, \bar \nu >0.
$$
\end{enumerate}

Numerical quadrature must be used for integral evaluation.
Alternatively the FFT algorithm can be used to efficiently
approximate integrals (see \cite{Lee04}) and then a standard
root-finding routine will find the required solutions.

\section{Regime-Switching Jump Diffusions and measure change}

Let us consider on a filtered probability space $(\Omega,
\mathcal{F}, \mathcal{F}_t, \P)$ a stochastic process of the form
$S_t=S_0\e^{X_{t}}$, $S_0 >0$, modeling the value, or price, of a
risky asset for $t \in [0,T]$. We consider a jump-diffusion
setting in which the jump process is described as a marked point
process (MPP), the parameters of which are driven by a finite
state and continuous time Markov chain.

We briefly recall here (see e.g. Runggaldier (2003)) that a MPP
can be characterized through the couple $(T_n, Y_n)$, where
$\{T_n\}$ is an univariate point process on $\R^+$ and $\{Y_n\}$
is a sequence of random variables on a given measurable space
$(E,{\cal E})$, as a random measure $p(dy,dt)$ for which
$$
\int_0^t \int_E \mathcal{H}(y,s) p(dy,ds) = \sum_{n=1}^{N_t}
\mathcal{H}(Y_n, T_n)
$$
$N_t$ being the Poisson point process. The corresponding intensity
for $p$ is a measure-valued process $\lambda_t(dy)$ for which
$$
\int_0^t \int_E \mathcal{H}(y,s) (p(dy,ds)- \lambda_s(dy)ds)
$$
is a martingale for each predictable process $\mathcal{H}$ and it
characterizes the MPP. A common form of the intensity is
$\lambda_t(dy) = \lambda_t m_t(dy)$, where $\lambda_t$ represents
the intensity of the Poisson counting process and $m_t(dy)$ is a
probability measure on the mark space $(E,\mathcal{E})$ describing
the jump component. Finally, the couple $(\lambda_t, m_t(dy))$ is
called the $(\P,\mathcal{F}_t)$-\emph{local characteristic} of
$p(ds,dy)$. This setting has been introduced in the financial
literature by Bjork et al. (1997). Although jump diffusion models
can be described in somewhat different ways, the approach based on
MPP turns out to be particularly useful for managing absolutely
continuous change of measures. As a matter of fact, for our
application we have to specify the dynamic model under both the
objective (or historical) measure $\P$ and an equivalent
risk-neutral (or pricing) measure $\Q$.

Let $\alpha(t)$ be a continuous time, homogeneous and stationary
Markov Chain on the state space $\mathcal{S} = \{1,2,\ldots,M \}$
with a generator $H \in \mathbb{R}^{M\times M}$; furthermore, $\mu
: \mathcal{S} \rightarrow \mathbb{R}$, $\sigma : \mathcal{S}
\rightarrow \mathbb{R}$ and $\gamma : E \times \mathcal{S}
\rightarrow \mathbb{R}$ are given functions, $(E,\mathcal{E})$
being the measurable mark space. Without loss of generality, we
can assume in the following $E \subseteq \mathbb{R}$. In a given
interval $0 \leq t \leq T$, we consider therefore the dynamic
\begin{equation}
d X(t)  =  (\mu(\alpha(t)) - \half \sigma^2(\alpha(t)) dt +
\sigma(\alpha(t)) d W(t) +  \int_E
\gamma(y,\alpha(t^-))p^{\alpha}(dy,dt), \ \ X(0)=0,
\end{equation}
where $W(t)$ is a standard brownian motion and $p^{\alpha}(dy,dt)$
is a MPP characterized by the intensity
$$
\lambda^{\alpha}_t(dy) \equiv \lambda(\alpha) m(\alpha,dy).
$$
Here $\lambda(\cdot)$ represents the (regime-switching) intensity
of the Poisson process $N_t$, while $m(\cdot,dy)$ are a set of
probability measures on $E$, one for each state (regime) $i \in
\mathcal{S}$ of the chain. The function $\gamma(y,\alpha)$
represents the jump amplitude relative to the mark $y$ in regime
$\alpha$. Throughout the paper we assume that the processes
$\alpha(\cdot)$ and $W(\cdot)$ are independent and that $W(\cdot)$
and $p^{\alpha}(dy,dt)$ are conditionally independent given
$\alpha(t)$. We denote $\mathcal{F}^{\alpha}_t = \sigma
\{\alpha(s): 0 \leq s \leq t \}$ the $\sigma$-algebra generated by
the Markov chain. Furthermore, we assume that
$\E[\e^{\gamma(Y,\alpha)}] = \int_E \e^{\gamma(y,\alpha)}
m(\alpha,dy)$ is finite for each regime $\alpha$. As usual, we
also define the compensated point process $q^{\alpha}(dy,dt) =
p^{\alpha}(dy,dt)-\lambda(\alpha(t^-)) m(\alpha(t^-),dy) dt$ in
such a way $\int_0^t \int_E H(y,\alpha(s^-))q^{\alpha}(dy,ds)$ is
a martingale in $t$ for each predictable process $H$ satisfying
appropriate integrability conditions.

An application of the generalized Ito's Formula gives the
corresponding jump-diffusion SDE for the asset price
\begin{eqnarray}\label{rsjdm_sde}
\frac{d S(t)}{S(t^-)} & = & \mu(\alpha(t)) dt + \sigma(\alpha(t))
dW(t) + \int_E (\e^{\gamma(y,\alpha(t-))}-1) p^{\alpha}(dy,dt) \\
 & = & [\mu(\alpha(t))+\lambda(\alpha(t))] \kappa(\alpha(t)))) dt + \sigma(\alpha(t)) dW(t) + \int_E
(\e^{\gamma(y,\alpha(t-))}-1) q^{\alpha}(dy,dt),  \nonumber
\end{eqnarray}
with $S(0)=s_0$ and $\kappa(\alpha) =
\E[\e^{\gamma(Y,\alpha)}-1]$.
%where we defined $g(y,\alpha) = \e^{\gamma(y,\alpha)}-1$.

\paragraph{Measure changes.}

An absolutely continuous transformation of measures in a
jump-diffusion setting allows to change the intensities of the MPP
and the Markov chain in addition to the translation of the Wiener
process (see Runggaldier (2003)). In this context it results
convenient to represent the underlying Markov chain itself as a
MPP (see Landen (2001)) in such a way
\begin{equation}
d\alpha(t)  =  \int_\mathrm{S} \delta (z) \upsilon(dt,dz)
\end{equation}
where $\upsilon(dt,dz)$ is a marked point process with  finite
mark space $(\mathrm{S},\mathcal{P}(\mathrm{S}) )$, $\mathrm{S} =
\{z=(i,j):i\neq j, i,j \in \{1,2, \ldots, N\}\}$ and
$\mathcal{P}(\mathrm{S}) = 2^\mathrm{S}$ and compensator
\begin{equation} \label{compensatore}
\lambda_{\upsilon}(t,\alpha(t-),dz)=\sum_{i \neq j} h_{ij}
1_{(\alpha(t-)=i)} \epsilon_{(i,j)}(dz)
\end{equation}
$\epsilon_{(i,j)}(dz)$ being the Dirac measure. In the previous
formula the numbers $h_{ij}$ are positive and such that
$\sum_{j\neq i,j=1}^Nh_{ij}=-h_{ii}$, for $i=1,\ldots,N$.

\bigskip

Consequently, let $\vartheta_t(\alpha)$ be a square integrable
predictable processes, $h_t(\alpha,y)$ a non-negative function
such that
$$
\int_E h_t(\alpha,y) m(\alpha,dy) = 1, \forall t \in [0,T]
$$
and let $\psi(\alpha)$ and $\Phi(z)$ be strictly positive
functions defined on $\mathcal{S}$ and $\mathrm{S}$, respectively.
We can define a new measure $\Q$ on the measurable space by
setting
\begin{equation} \label{meas_change}
\left \{ \begin{array}{l}
          dW^Q_t =  dW_t - \vartheta_t(\alpha(t)) dt \\
          q^{Q,\alpha}(dt,dy) = p(dt,dy) -\psi(\alpha(t)) \lambda(\alpha(t)) h_t(\alpha(t),y)
           m(\alpha(t),dy) dt \\
           \lambda_{\upsilon}^Q(t,\alpha,dz) =
           \Phi(z) \lambda_{\upsilon}(t,\alpha,dz).
          \end{array} \right.
\end{equation}
Besides the translation of the Wiener process $W_t$, we perform a
change in the intensity of the MPP giving the compensated process
$q^{Q,\alpha}(dt,dy)$ with $(\Q, \mathcal{F}_t)$-local
characteristic $(\psi(\alpha(t)) \lambda(\alpha(t))$,
$h_t(\alpha(t),y) m(\alpha(t),dy))$ and a change of the intensity
of the Markov chain which under $\Q$ has generator $H^Q =
\{h_{ij}^Q \}$ where
$$
h_{ij}^Q = \Phi(i,j) h_{ij}, \ \ \ h_{ii}^Q = - \sum_{k=1, k \neq
i}^N \Phi(i,k) h_{ik}, \ \ \ \ i=1,\ldots, N.
$$

By taking the Radon-Nikodym derivative
\begin{eqnarray} \label{rad_nycod}
L_t & = & \exp \left(- \half \int_0^t \vartheta_s(\alpha(s))^2
ds + \int_0^t \vartheta_s(\alpha(s)) dW^Q_s  \nonumber \right.\\
    &  + &   \int_0^t (1-\psi(\alpha(s)) \lambda(\alpha(s)) ds + \int_0^t \int_E
\log(\psi(\alpha(s)) h_t(\alpha(s),y)) p(ds,dy)  \\
    & + & \left. \int_0^t (1-\Phi(z)) \lambda_{\upsilon}(t,\alpha(t-),dz) ds
          + \int_0^t \int_\mathrm{S} \log(\Phi(z))
          \upsilon(ds,dz)\right) \nonumber
\end{eqnarray}
and supposing that $\E^{\P}[L_t] = 1$ for all $t$, we have a
probability measure $\Q$ on $\mathcal{F}$ equivalent to $\P$ with
$d\Q = L_T d\P$, under which
\begin{eqnarray}
\frac{d S(t)}{S(t^-)} & = & \left[
\mu(\alpha(t))+\sigma(\alpha(t)) \vartheta_t(\alpha(t)) +
\psi(\alpha(t)) \lambda(\alpha(t)) \kappa^Q(\alpha(t)) \right] dt
+ \sigma(\alpha(t)) dW^Q(t) \nonumber \\  & + & \int_E
(\e^{\gamma(y,\alpha(t-))}-1) q^{Q,\alpha}(dy,dt),
\end{eqnarray}
where $\kappa^Q(\alpha) = \E^Q[\e^{\gamma(Y,\alpha)}-1]$.

In order to price derivatives under the model (\ref{rsjdm_sde}) we
need to specify a risk-neutral or martingale measure, that is a
measure under which the discounted price process $\e^{-r t} S_t$
is a martingale. This is done by taking
\begin{equation} \label{thetarisk}
\vartheta(\alpha) \equiv \frac{r - \mu(\alpha) - \psi(\alpha)
\lambda(\alpha) \kappa^Q(\alpha)}{\sigma(\alpha)}
\end{equation}
from which we finally get the risk-neutral dynamic for the
underlying
\begin{equation} \label{rsjdQ_sde}
\frac{d S(t)}{S(t^-)}  = r dt + \sigma(\alpha(t)) dW^Q(t) + \int_E
(\e^{\gamma(y,\alpha(t-))}-1) q^{Q,\alpha}(dy,dt)
\end{equation}
$$
=  [r - \psi(\alpha(t)) \lambda(\alpha(t)) \kappa^Q(\alpha(t))] dt
+ \sigma(\alpha(t)) dW^Q(t) + \int_E (\e^{\gamma(y,\alpha(t-))}-1)
p^{Q,\alpha}(dy,dt).
$$

Consequently, from (\ref{thetarisk}) and (\ref{rsjdm_sde}), the
market price of risk is (see \cite{Runga})
$$
\rho(\alpha) \equiv \mu(\alpha) - \lambda(\alpha) \kappa(\alpha) -
r = \lambda(\alpha) [ \kappa(\alpha) - \psi(\alpha)
\kappa^Q(\alpha)] - \sigma(\alpha) \theta(\alpha)
$$
\begin{equation}
= \lambda(\alpha) \left [ \int_E
(e^{\gamma(y,\alpha)}-1)(1-\psi(\alpha) h_t(\alpha,y))
m(\alpha,dy) \right] - \sigma(\alpha) \theta(\alpha).
\end{equation}

Correspondingly, for the process $X(t)$ we have
\begin{eqnarray}
dX(t) & = & [r - \half \sigma^2(\alpha(t))] dt + \sigma(\alpha(t))
dW^Q(t) + \int_E \gamma(y,\alpha(t^-)) q^{Q,\alpha}(dy,dt)
\nonumber \\ & = & [r - \half \sigma^2(\alpha(t)) -
\psi(\alpha(t)) \lambda(\alpha(t))
\kappa^Q(\alpha(t)) ] dt + \sigma(\alpha(t)) dW^Q(t) \nonumber \\
& + & \int_E \gamma(y,\alpha(t^-)) p^{Q,\alpha}(dy,dt).
\label{rsjdXQ_sde}
\end{eqnarray}

The measure transformation defined by (\ref{meas_change}) through
(\ref{rad_nycod}) preserves the probability structure of the
stochastic process $X(t)$ under both $\mathcal{P}$ and
$\mathcal{Q}$. It worth noting that we can specify infinitely many
equivalent measures $\mathcal{Q}$. In practice, the usual way to
select one of the equivalent measures is to calibrate the model to
a set of observed data.

\paragraph{GFT for regime-switching jump-diffusions.}
%%%%%%%%%%%%%%%%%%%%%%%%%%%%%%%%%%%%%%%%%%%%%%%%%%%%%%%%%%%%%%%%%%%%%%%
%TO BE DISCUSSED: IS THE CHARACTERISTIC FUNCTION OF THE PROCESS
%X(t) INTEGRABLE? THIS IMPLIES THAT THE DISTRIBUTION FUNCTION IS
%ABSOLUTELY CONTINUOUS, I.E. DENSITY EXISTS, AND THE VAR PROBLEM IS
%EASIER.
%%%%%%%%%%%%%%%%%%%%%%%%%%%%%%%%%%%%%%%%%%%%%%%%%%%%%%%%%%%%%%%%%%%%%%%
The formulas (\ref{put_formula}) and (\ref{proba_formula}) depend
from the dynamic model only through its generalized Fourier
transform $\phi_X(z)$. In this section we report the GTF for the
regime-switching jump-diffusion model. Since we have to consider
the process $X_T$ under two different measure, we derive its
characteristic function for the following general dynamic
$$
dX(t) = \xi(\alpha(t)) dt + \sigma(\alpha(t)) dW(t) + \int_E
\gamma(y,\alpha(t^-)) p(dy,dt)
$$
where
\begin{equation} \label{xi_X}
\xi(\alpha) = \left\{ \begin{array}{cc}
                        r - \half \sigma^2(\alpha) -
\psi(\alpha) \lambda(\alpha) \kappa^Q(\alpha) & \mbox{under the measure} \ \ \Q \\
                        \mu(\alpha) - \half \sigma^2(\alpha) & \mbox{under the measure} \ \ \P \\
                      \end{array} \right.
\end{equation}
the MPP $p(dy,dt)$ has intensity
$$
\lambda(\alpha,dy) =  \left\{ \begin{array}{cc} \psi(\alpha)
\lambda(\alpha) h(\alpha,y) m(\alpha,dy) & \mbox{under the measure} \ \ \Q \\
\lambda(\alpha) m(\alpha,dy) & \mbox{under the measure} \ \ \P
\end{array} \right.
$$
and the Markov chain has generator $Q=\{q_{ij} \}_{i,j=1,\ldots,
N}$ where
\begin{equation} \label{qgen}
q_{ij} =  \left\{ \begin{array}{cc} \e^{\Phi(i,j)} h_{ij} & \mbox{under the measure} \ \ \Q \\
                               h_{ij}  & \mbox{under the measure} \ \ \P
\end{array} \right. \ \ \ i,j=1,\ldots, N, i \neq j, \ \mbox{and} \ q_{ii} =
- \sum_{k=1, k \neq i}^N q_{ik}.
\end{equation}

In \cite{Ram10} (see also \cite{Cho05}) it was proved the
following

{\proposition \label{propCFX} Let $\phi_j(z)=\E[\e^{\ci z
\gamma(Y(j),j)}]$ be the generalized Fourier transform of the jump
magnitude under the given measure. Then, by letting
\begin{equation}\label{theta1}
\vartheta_j(z) = z \xi(j)+\frac{1}{2} \ci z^2 \sigma^2(j) - \ci
\lambda(j)(\phi_j(z)-1)
\end{equation}
and $\tilde \vartheta_i(z) = \vartheta_j(z) - \vartheta_M(z)$, we
have
\begin{equation} \label{charfun2}
\begin{array}{lll}
\varphi_T(z) & = & \e^{\ci \vartheta_M(z) T} \left(\mathbf{1}'
\cdot \e^{(Q' + \ci \ \mathrm{diag}(\tilde \vartheta_1(z), \ldots,
\tilde \vartheta_{M-1}(z),0))T} \cdot \mathbb{I}(0) \right) \\ \\
 & = & \mathbf{1}' \cdot \e^{(Q' + \ci \
\mathrm{diag}(\vartheta_1(z), \ldots, \vartheta_{M}(z)))T} \cdot
\mathbb{I}(0), \end{array}
\end{equation}
where $\mathbf{1} = (1,\ldots,1)' \in \mathbb{R}^{M \times 1}$,
$\mathbb{I}(0) = (\mathbb{I}_{\alpha(0)=1}, \ldots,
\mathbb{I}_{\alpha(0)=M})'\in \mathbb{R}^{M \times 1}$ and $Q'$ is
the transpose of $Q$. }

\bigskip

Different models can be recovered with simple linear constraints
on the full parameter set of our dynamics (\ref{rsjdm_sde}),
(\ref{rsjdQ_sde}). This follows by noticing that if $\xi(i)=\xi,
\sigma(i)=\sigma$, $\lambda(i)=\lambda$ and $\phi_i(z) =\phi(z)$
we are implicitly assuming a unique regime so recovering the
well-known characteristic function of the (single-regime)
jump-diffusion dynamic $\varphi_T(z) = \exp( z \xi+\frac{1}{2} \ci
z^2 \sigma^2 - \ci \lambda(\phi(z)-1))$ which includes the
standard geometrical Brownian motion (GBM) ($\lambda=0$) and the
Merton jump-diffusion models (JDM). By letting $\lambda_i =0$ in
(\ref{charfun2}) we get the regime-switching version of GBM
(RSGBM) and finally the regime-switching jump diffusion model
(RSJDM) with the full set of parameters
$$
\xi_i, \sigma_i, \lambda_i, h_{ij}, \ \ \ \  i,j=1,\ldots,M.
$$

The evaluation of the characteristic function requires to compute
matrix exponentials for which efficient numerical techniques are
available (see Higham (2009)); conversely, the case $M=2$ can be
considered explicitly. The following can be proved (see
\cite{Ram10} and the references therein).

{\proposition \label{twocharfun} Let $y_{1,2}$ be the solutions of
the quadratic equation $y^2+(q_1+q_2-\ci \theta) y - \ci \theta
q_2 = 0$ and
$$
\begin{array}{lll}
\mathrm{q}_1^{T}(\theta) & = & \frac{1}{y_1-y_2} \left( \e^{y_1
T}(y_1+q_1+q_2)-\e^{y_2 T} (y_2+q_1+q_2)\right) \\ \\
\mathrm{q}_2^{T}(\theta) & = & \frac{1}{y_1-y_2} \left(\e^{y_1
T}(y_1+q_1+q_2-\ci \theta)-\e^{y_2 T} (y_2+q_1+q_2-\ci \theta)
\right).
\end{array}
$$

Then
$$
\E_t[\e^{\ci \theta T_1}] = \mathbb{I}_{\alpha(t)=1}
\mathrm{q}_1^{T}(\theta)  +  \mathbb{I}_{\alpha(t)=2}
\mathrm{q}_2^{T}(\theta)
$$
and therefore
$$
\varphi_X(z) = \e^{\ci \vartheta_2(z) T} \left(
\mathbb{I}_{\alpha(t)=1} \mathrm{q}_1^{T}(\theta(z)) +
\mathbb{I}_{\alpha(t)=2} \mathrm{q}_2^{T}(\theta(z))  \right)
$$

$ \Box$ }

\section{Numerical results}

In this section we report results about two kind of numerical
experiments:

\begin{enumerate}
\item calculation of quantiles for RSJD. The objective is to study
the behavior of such a quantities by varying diffusions and jumps
parameters.
\item valuation of the optimal hedging strategy in the regime-switching
jump-diffusion framework.
\end{enumerate}

We consider a two-state regime switching version of the
jump-diffusion model with gaussian jumps. This is defined by
choosing $\gamma(y,\alpha)=y$ and two kinds of normal jumps, i.e.
$Y(i) \sim \mathcal{N}(a_i,b_i)$ from which $\kappa(i)=
\E[(\e^{Y(i)}-1)] = \e^{a_i + b_i^2/2}-1$, $i=1,2$. The two state
Markov chain $\alpha(t) \in \mathcal{S} = \{1,2 \}$ has
generator under the chosen measure $Q = \left(%
\begin{array}{cc}
  -q_1 & q_1 \\
  q_2 & -q_2 \\
\end{array}%
\right)$. Let $\sigma_i, \lambda_i > 0$ and $\mu_i, i=1,2$ be
given parameters: the regime switching jump-diffusion Merton model
is defined as
\begin{equation} \label{merton1}
dX(t) = \xi(\alpha(t)) dt + \sigma(\alpha(t)) dW(t) + \int_0^t
\int_E  y p^{\alpha}(dy,ds)
\end{equation}
where $\xi(\alpha(t))$ is given by (\ref{xi_X}),
$\lambda(t,\alpha(t),dy) = \lambda(\alpha(t)) \phi_{\alpha(t)}(y)
dy$ is the intensity process of the Poisson jump component and
$\lambda(\alpha(t)) \in \{\lambda_1,\lambda_2 \}$,
$\sigma(\alpha(t)) \in \{\sigma_1,\sigma_2 \}$, $\mu(\alpha(t))
\in \{\mu_1,\mu_2 \}$, $\phi_{i}(y)$ being the density of a normal
distribution $\mathcal{N}(a_i, b_i)$, $i=1,2$.

All numerical procedures were implemented in the
MatLab$^\copyright$ framework.

\subsection{Quantiles and VaR Calculation}

Quantiles were computed by solving the equation
$$
\frac{(q/S_0)^{\nu}}{\pi} \Re \left( \int_{0}^{+\infty} \e^{- \ci
u \log(q/S_0)} \frac{\phi_X^{\mathcal{P}}(u + \ci \nu)}{\nu-\ci u}
du \right) = \alpha,
$$
with respect to $q$ (see Section 3). A standard root-search
algorithm was used together with the Gauss-Lobatto quadrature for
approximating the integral. Few milliseconds were needed to get
the required quantile on an Intel$^\copyright$ Core i5.

\begin{figure}
\begin{center}
\includegraphics[width=12cm,height=7cm]{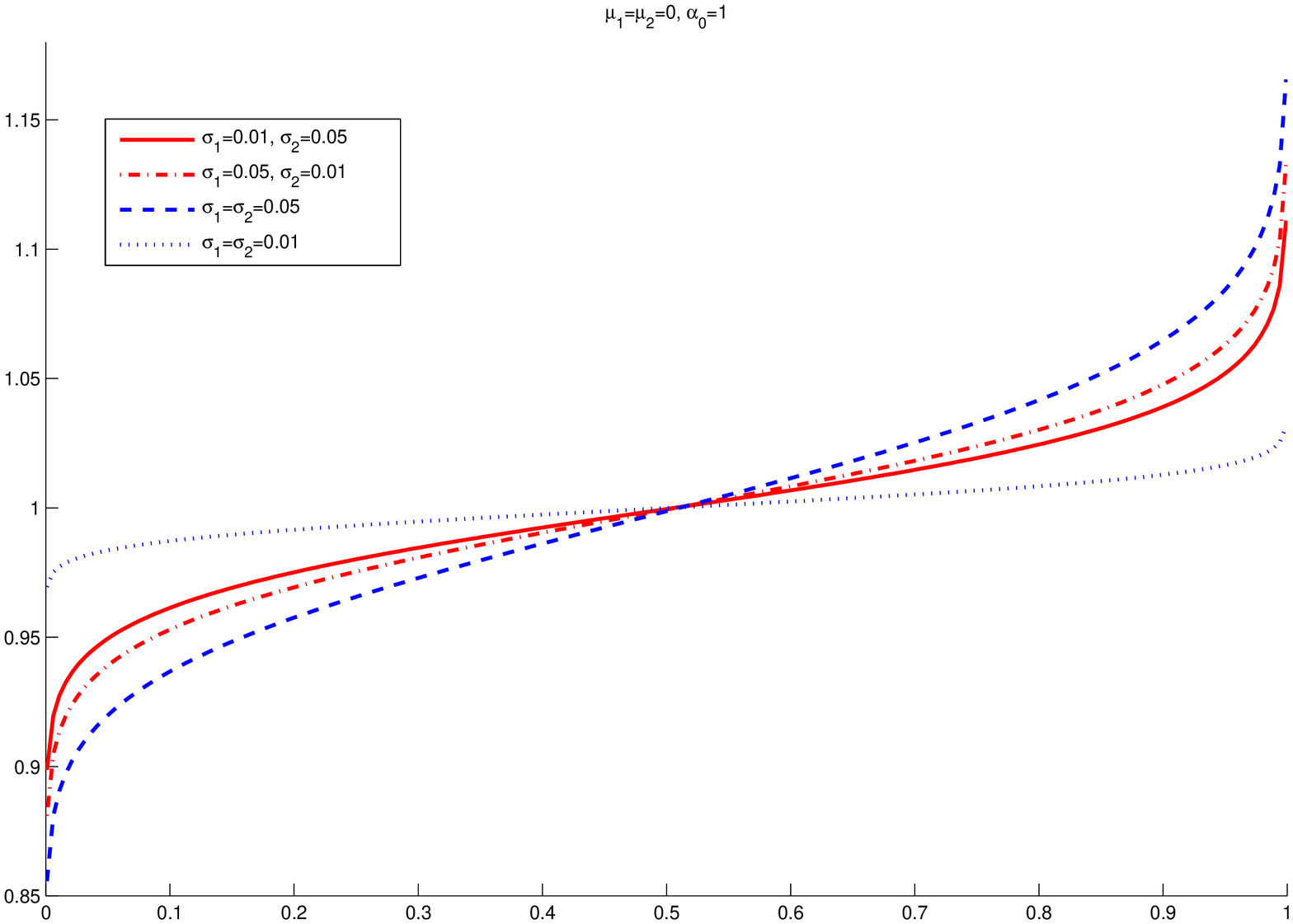}
\caption{\small Regime-switching GBM, starting state $\alpha_0=1$,
$q_1=q_2=5$ and $T=1$. Comparison with standard GBM. Analogous
results for starting state $\alpha_0=2$.}
\label{RSGBM_Psigmamix1Fig}
\end{center}
\end{figure}

\begin{figure}
\begin{center}
\includegraphics[width=12cm,height=7cm]{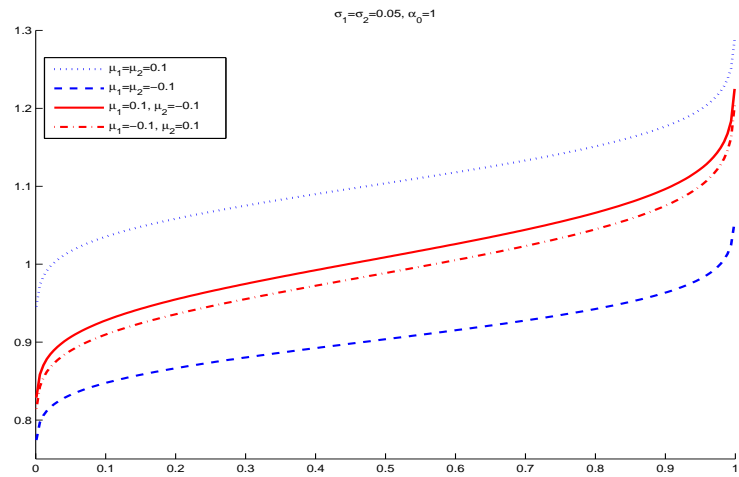}
\caption{\small Regime-switching GBM, starting state $\alpha_0=1$,
$q_1=q_2=5$ and $T=1$.} \label{RSGBM_Pmumix1Fig}
\end{center}
\end{figure}

\begin{figure}
\begin{center}
\includegraphics[width=12cm,height=7cm]{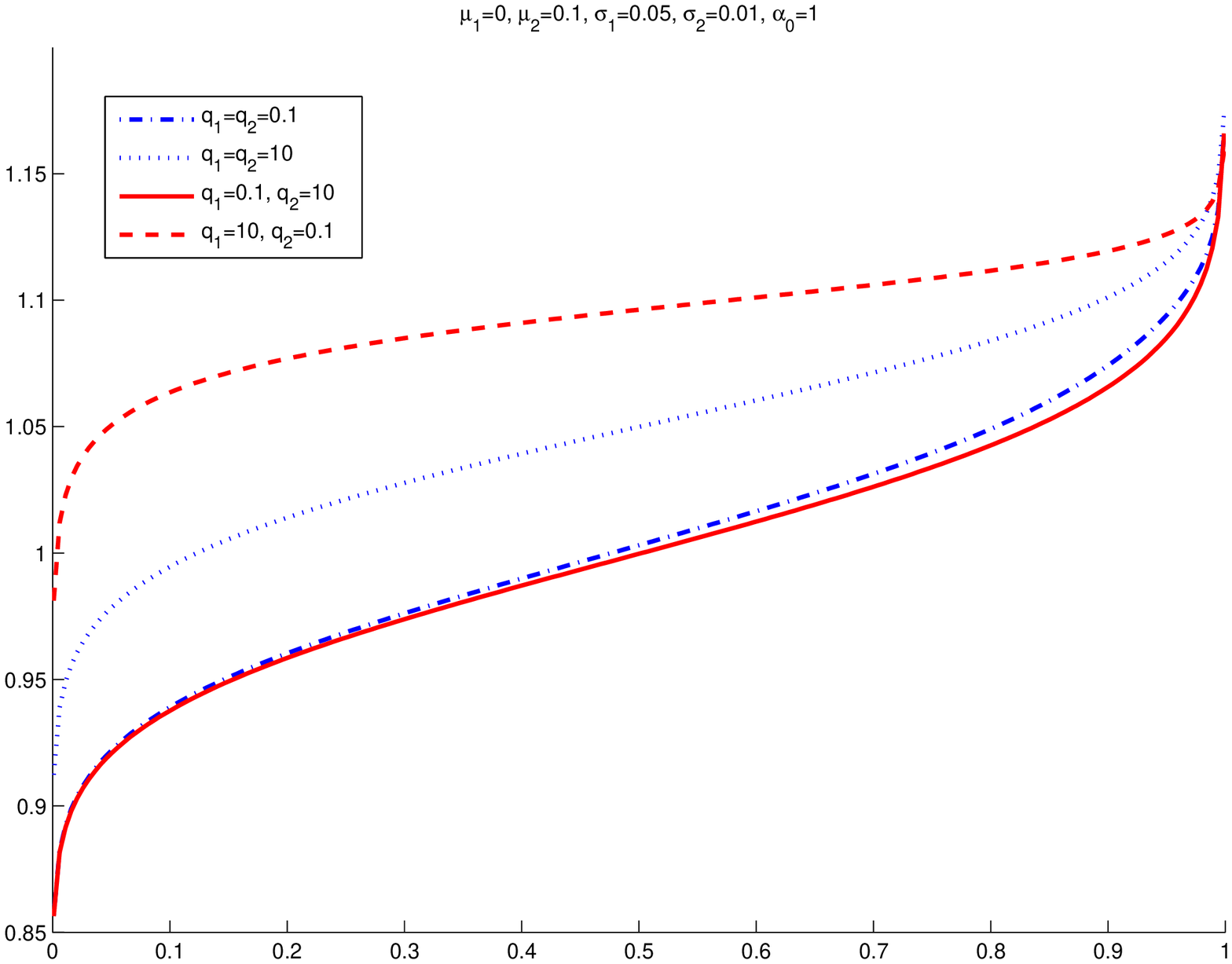}
\caption{\small Regime-switching GBM for different values of $q_i$
and $T=1$.} \label{RSGBM_Pq1q2Fig}
\end{center}
\end{figure}

\begin{figure}
\begin{center}
\includegraphics[width=12cm,height=7cm]{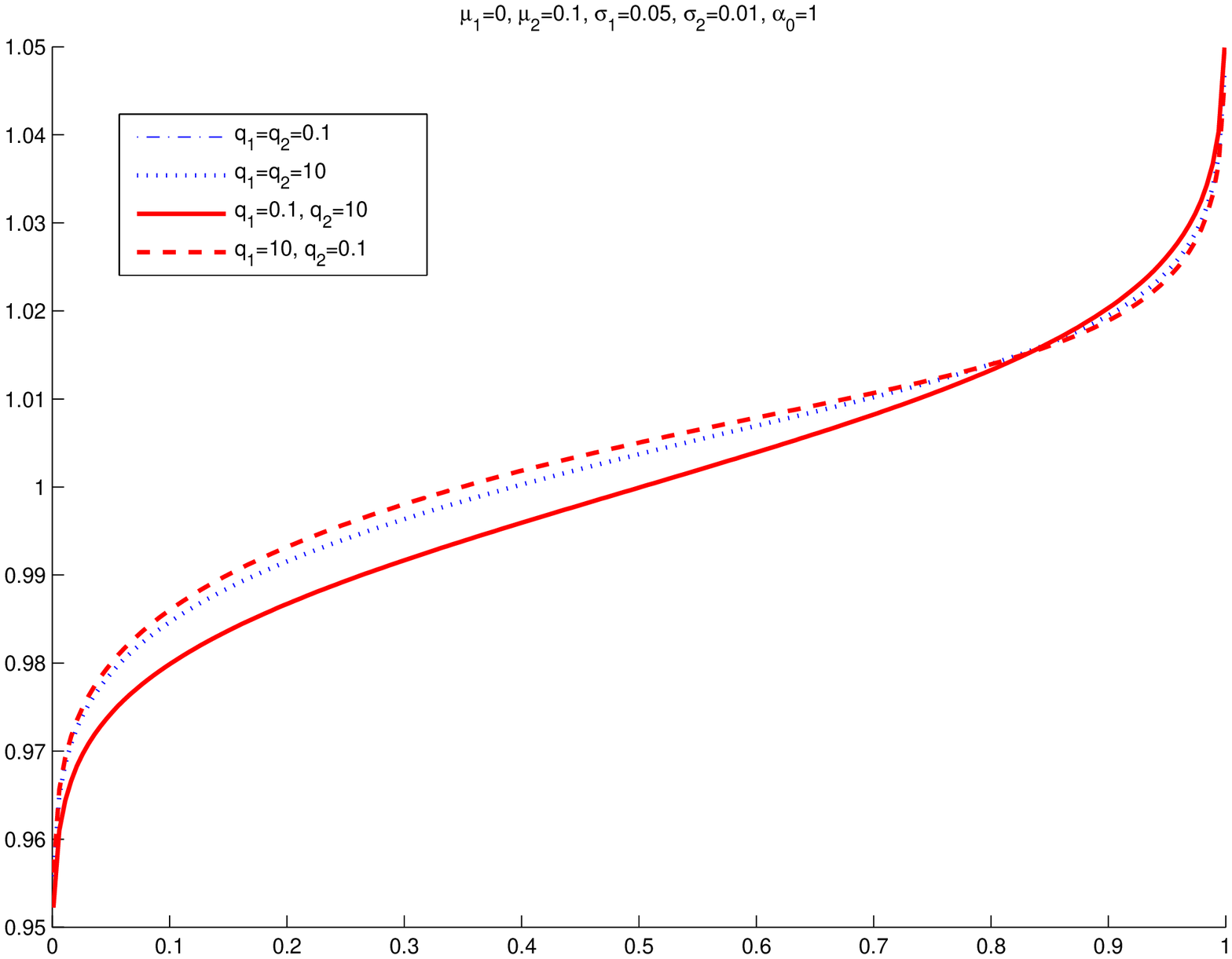}
\caption{\small Regime-switching GBM for different values of $q_i$
and $T=10$.} \label{RSGBM_Pq1q22Fig}
\end{center}
\end{figure}

We calculate the quantiles of the RSJD models for different values
of the parameters under the historical probability $\P$. We start
by considering the simple regime-switching geometrical brownian
motion ($\lambda_i=a_i=b_i=0$): as expected, the effect of
switching drift and volatility results in a sort of mixing
behavior between the corresponding GBM models, see figures
(\ref{RSGBM_Psigmamix1Fig}), (\ref{RSGBM_Pmumix1Fig}), and such
effect becomes more evident for growing time $T$. The frequency of
Markov chain switching, driven by the generator $Q$, induces a
quite different behavior, whose impact also depends on the value
of $T$ (figures (\ref{RSGBM_Pq1q2Fig}), (\ref{RSGBM_Pq1q22Fig}).

\begin{figure}[h]
\begin{center}
\includegraphics[width=12cm,height=7cm]{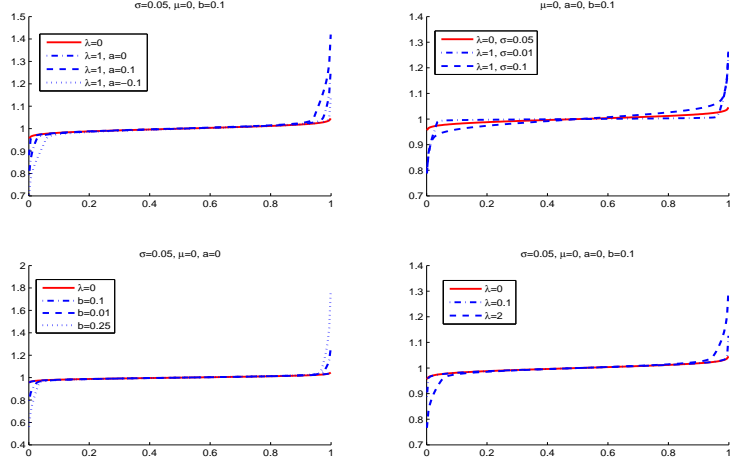}
\caption{\small Jump diffusion model with gaussian jumps: red
lines correspond to GBM.  $T=1$.} \label{JDM_Pall1Fig}
\end{center}
\end{figure}

It is well-known (\cite{CT}) that introducing (gaussian) jumps in
the GBM dynamic has a great impact on the tails of the
distribution, see figure (\ref{JDM_Pall1Fig}). As before, the
Markov switching generates a mixing effect: quantiles curves of
the RSJD models are between the curves of the corresponding JD
models without regime switching.

\subsection{Optimal risk management}
In the second set of numerical experiments we face two main
issues: (a) Risk Reduction: how much reduction of risk is obtained
by implementing the optimal hedging strategy in the RSJD framework
and in particular what is the impact of jumps and switching
regimes on the main quantities driving the strategy? (b)
Misspecified Modeling: what is the effect of a wrong model
specification which discards regime switchings and jumps, when
they are indeed present in the market, and consider the simpler
GBM model?

\paragraph{(a) Risk reduction and model impact.}

\begin{figure}
\begin{center}
\includegraphics[width=12cm,height=8cm]{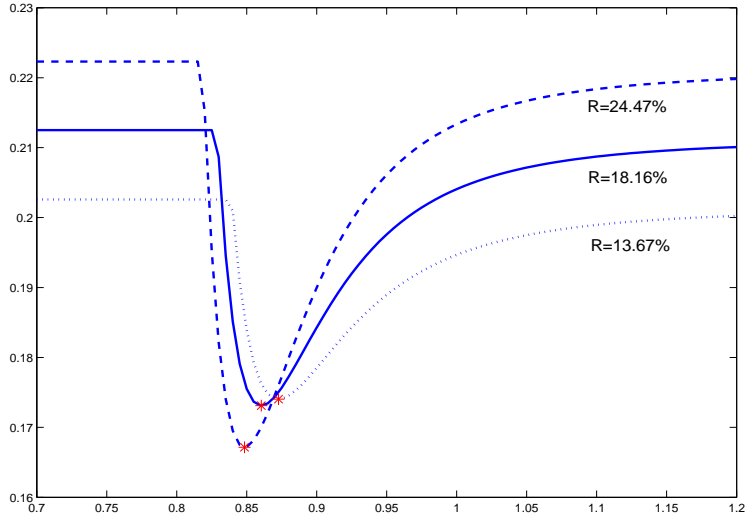}
\caption{\small Risk reduction in GBM model: the continuous line
corresponds to a model in which the risk is not priced, the dotted
line to a variation of $+25\%$ of the drift and the dashed line to
a variation of $-25\%$. The other parameters are $\sigma=0.1$,
$r=0.05$, $T=1$ and $S_0=1$. } \label{GBM_Red1Fig}
\end{center}
\end{figure}

\begin{figure}
\begin{center}
\includegraphics[width=7.5cm,height=6cm]{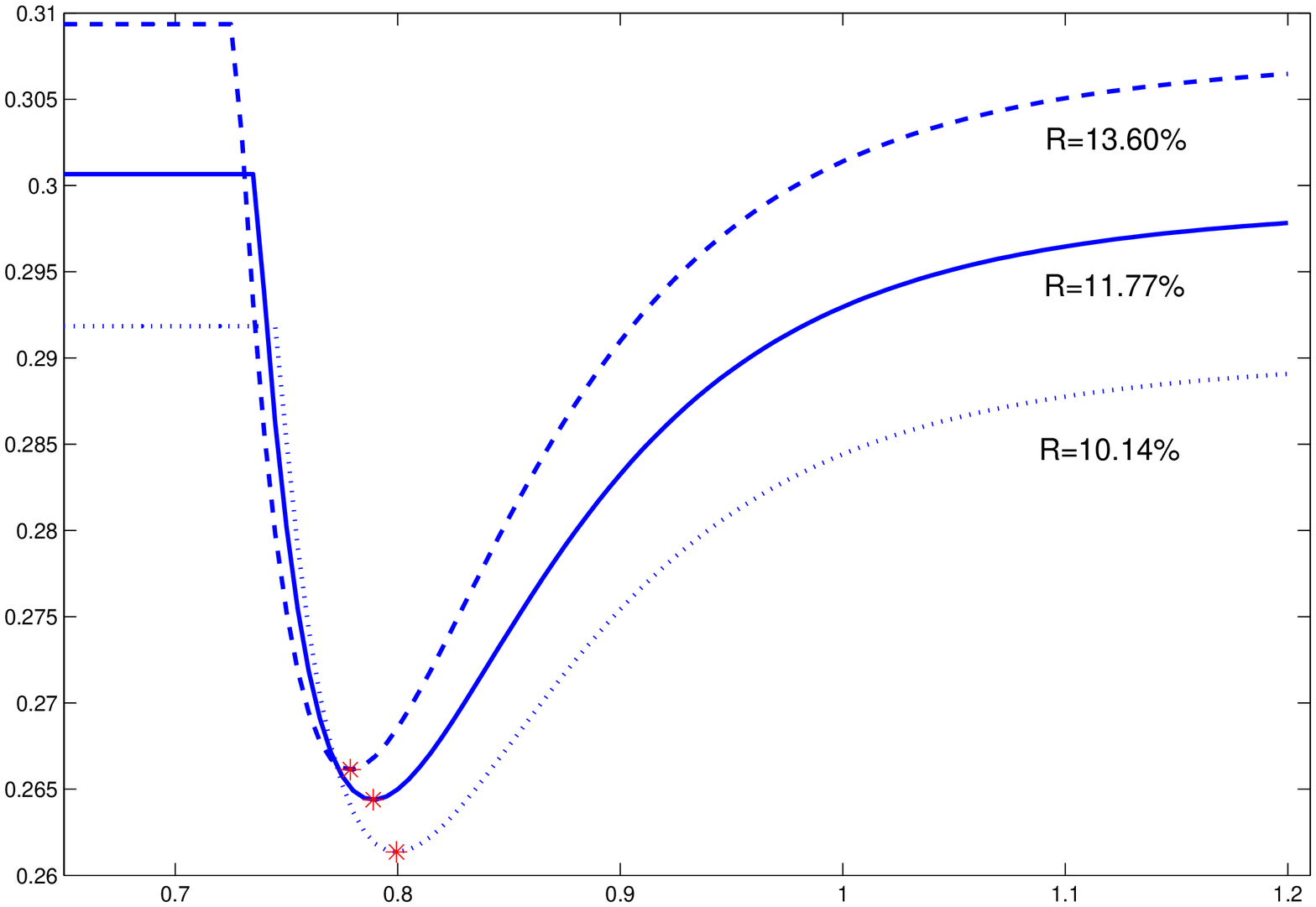}
\includegraphics[width=7.5cm,height=6cm]{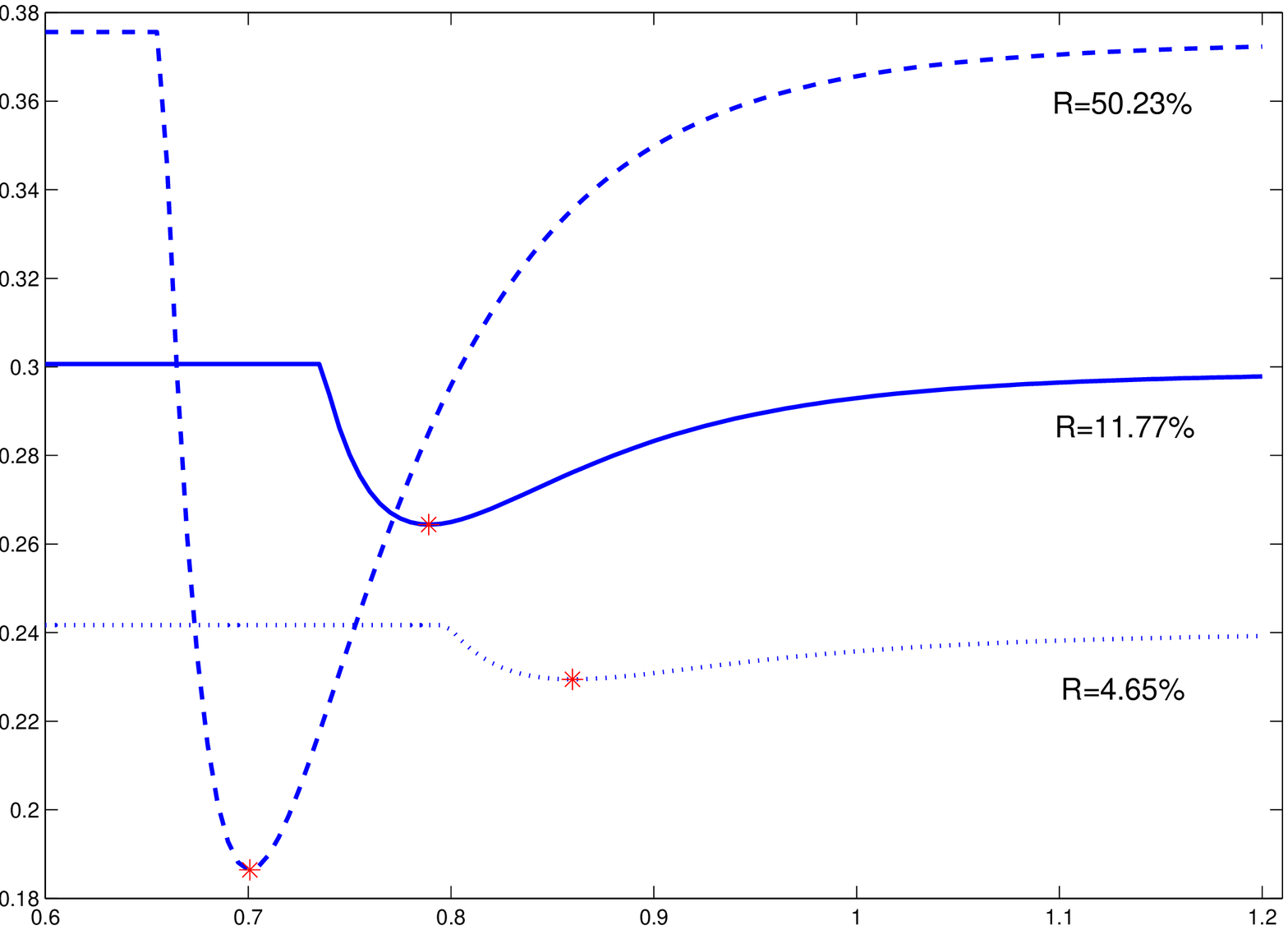}
\caption{\small Risk reduction in JDM model: the continuous line
corresponds to a model in which the risk is not priced, the dotted
line to a variation of $+25\%$ of the drift ($\mu$ - left figure)
and mean jump ($a$ - right figure) and the dashed line to a
variation of $-25\%$. The other parameters are $\sigma=0.1$,
$r=0.05$, $T=1$ and $S_0=1$. } \label{JDM_Red_mu1Fig}
\end{center}
\end{figure}

\begin{figure}
\begin{center}
\includegraphics[width=7.5cm,height=6cm]{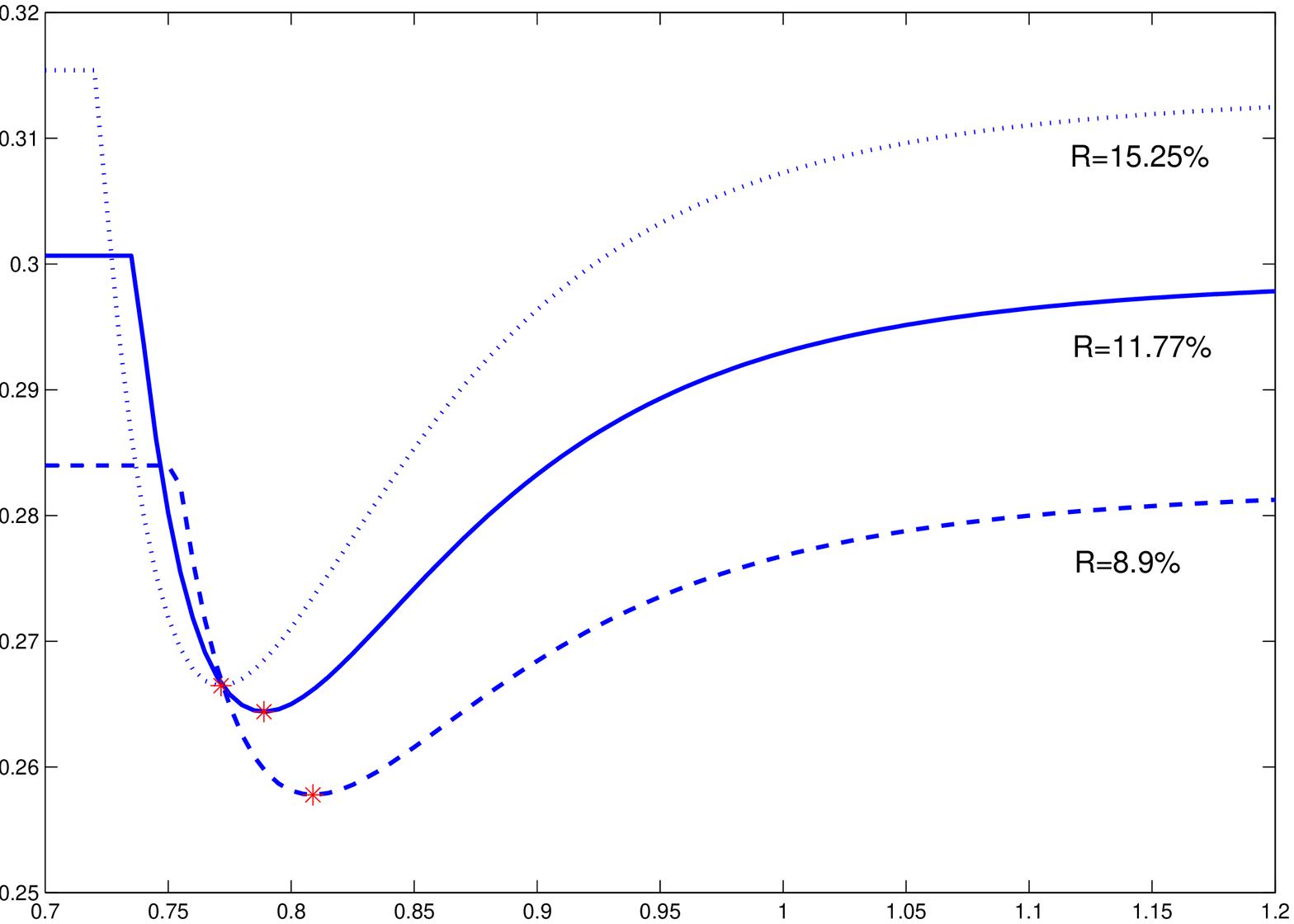}
\includegraphics[width=7.5cm,height=6cm]{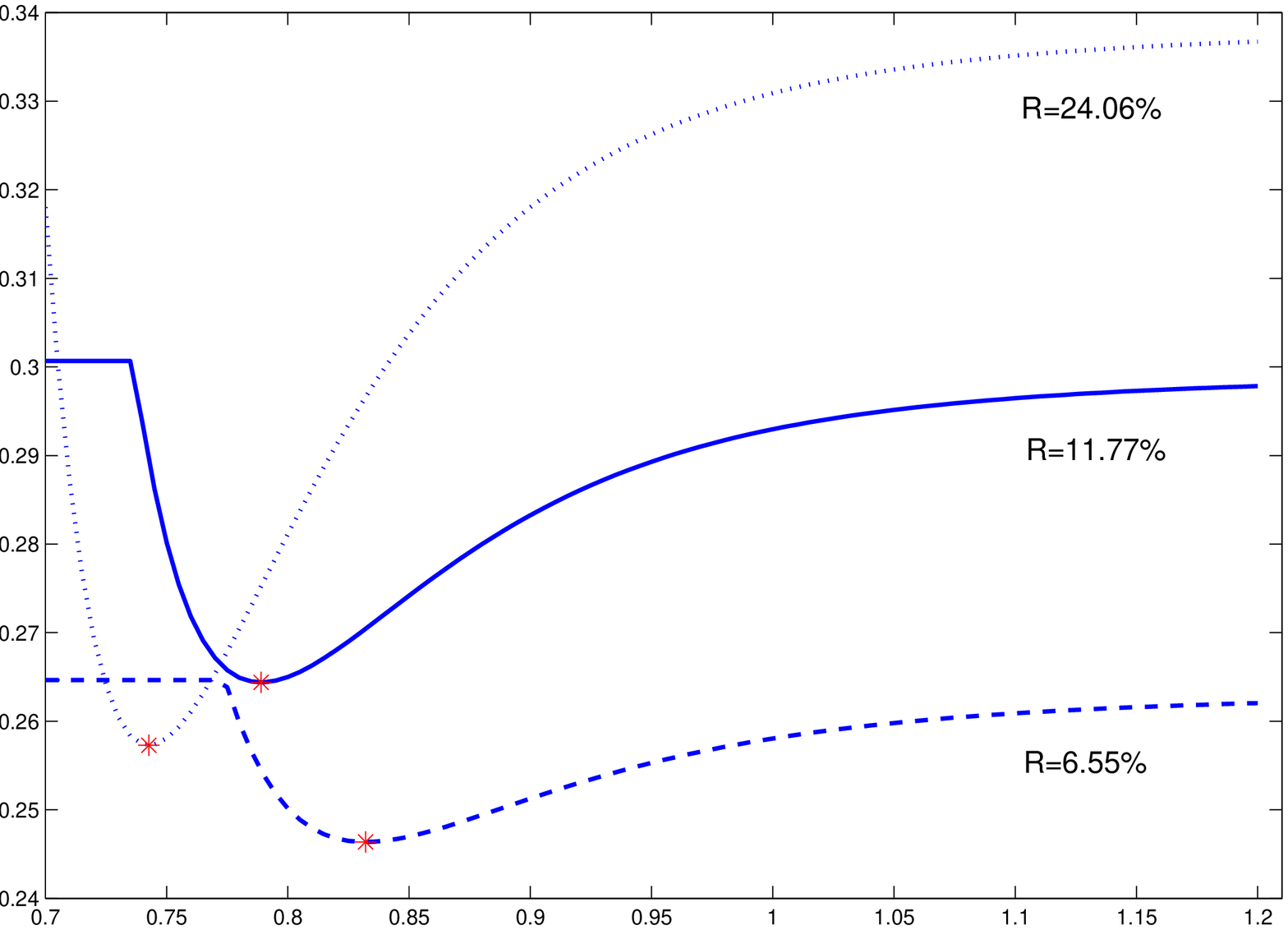}
\caption{\small Risk reduction in JDM model: the continuous line
corresponds to a model in which the risk is not priced, the dotted
line to a variation of $+25\%$ of the jump intensity ($\lambda$ -
left figure) and jump std ($b$ - right figure) and the dashed line
to a variation of $-25\%$. The other parameters are $\sigma=0.1$,
$r=0.05$, $\lambda=1$, $a=0$, $T=1$ and $S_0=1$. }
\label{JDM_Red_lam1Fig}
\end{center}
\end{figure}

\begin{figure}
\begin{center}
\includegraphics[width=7.5cm,height=6cm]{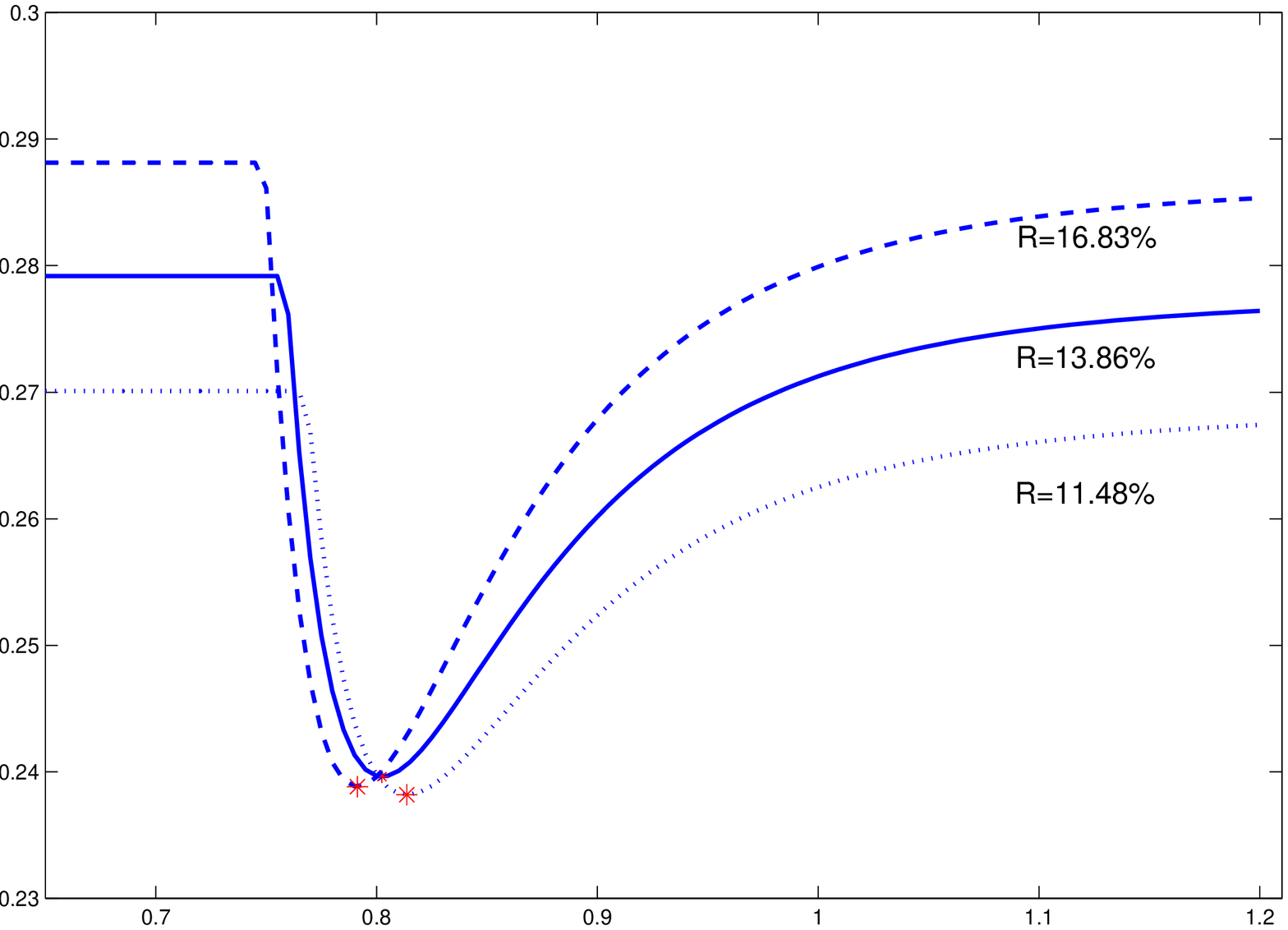}
\includegraphics[width=7.5cm,height=6cm]{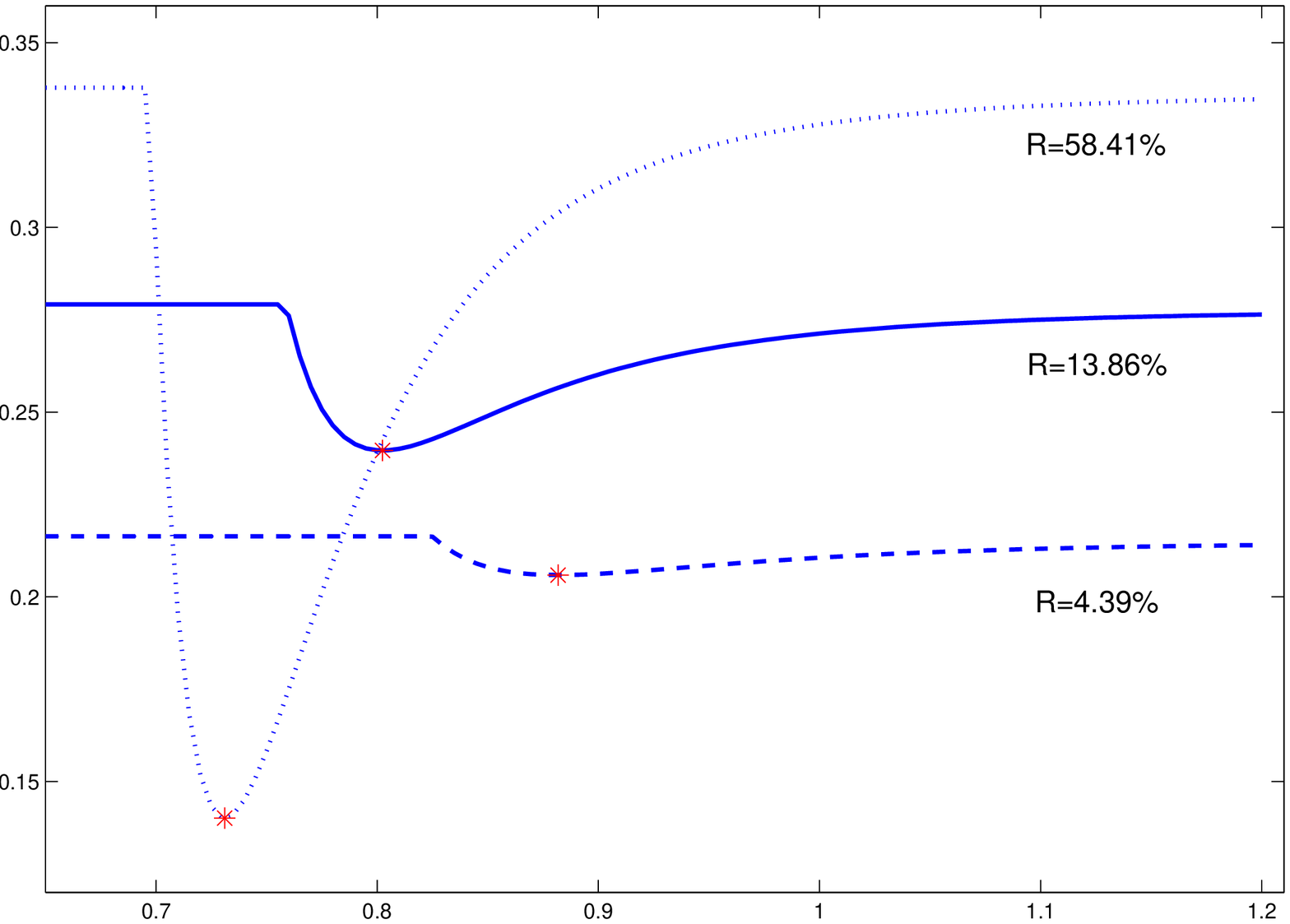}
\caption{\small Risk reduction in RS model: the continuous line
corresponds to a model in which the risk is not priced, the dotted
line to a variation of $+25\%$ of the drift ($\mu_1, \mu_2$ - left
figure) and volatility ($\sigma_1, \sigma_2$ - right figure) and
the dashed line to a variation of $-25\%$. The parameters are
$\sigma_1=0.15, \sigma_2=0.05$, $r=0.05$, $q_1=q_2=1$, $T=1$ and
$S_0=1$. No substantial changes are observed for this kind of
variation of the parameters $q_i$.} \label{RS_Red1Fig}
\end{center}
\end{figure}

\begin{figure}
\begin{center}
\includegraphics[width=7.5cm,height=6cm]{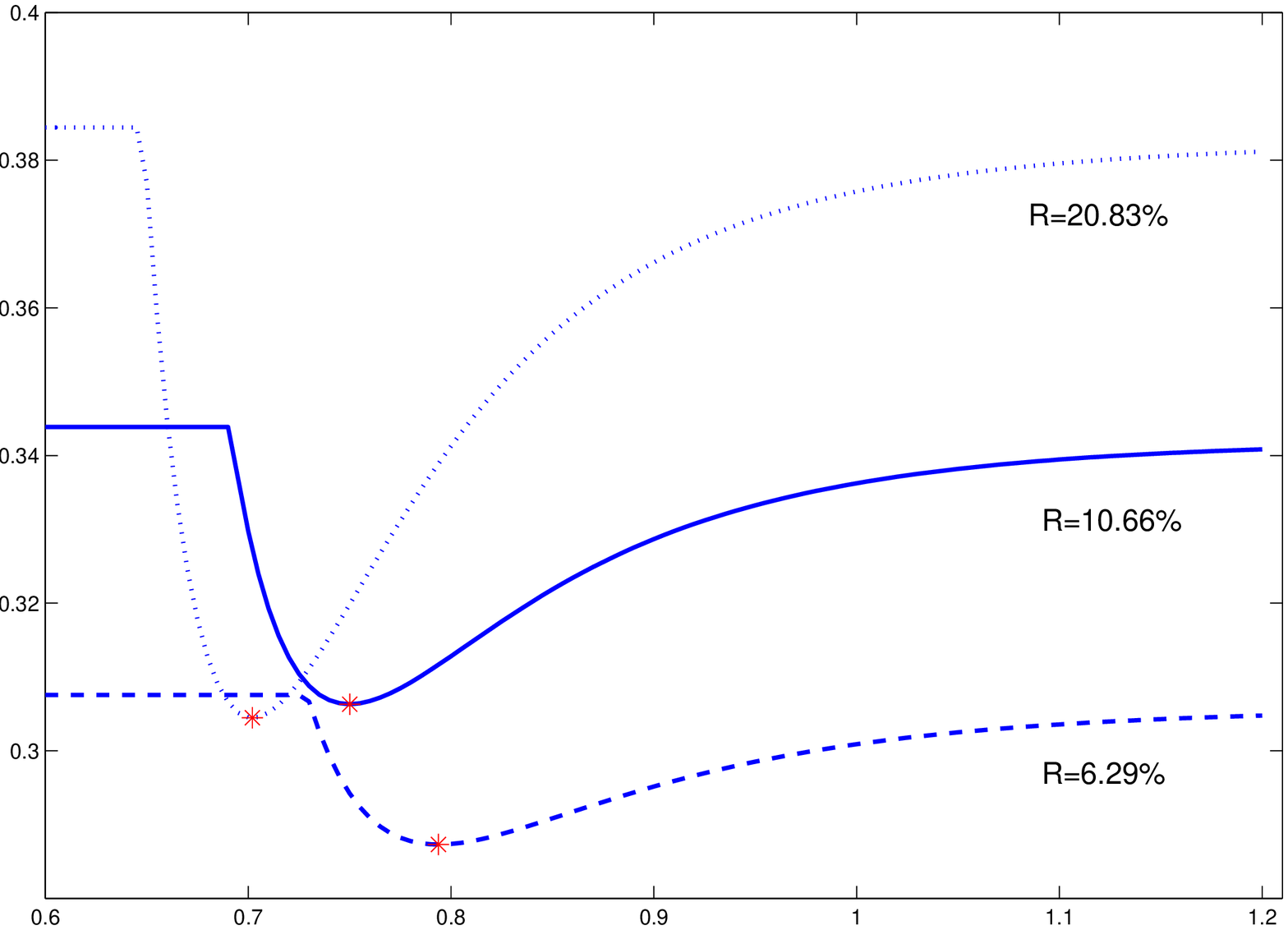}
\includegraphics[width=7.5cm,height=6cm]{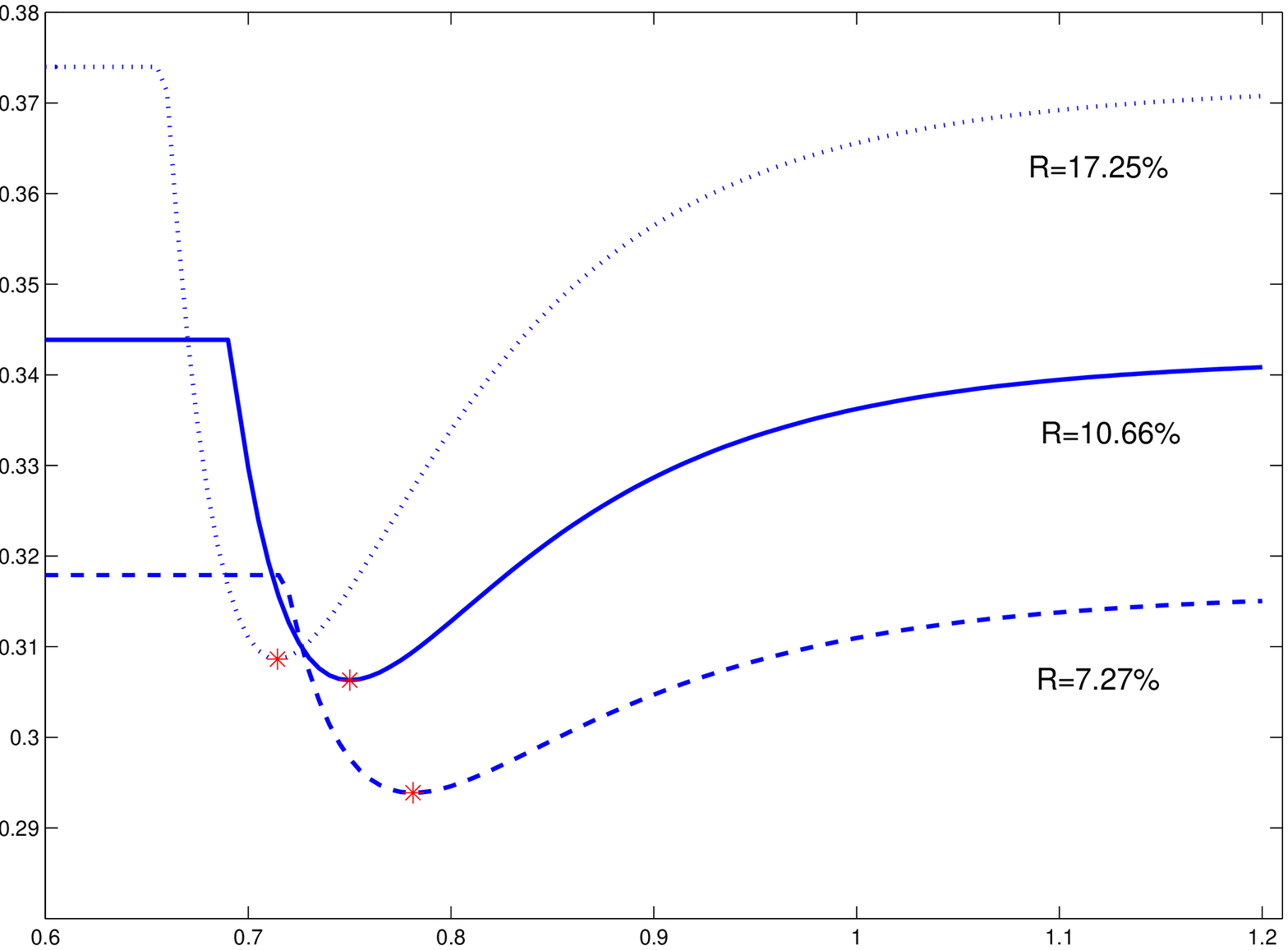}
\caption{\small Risk reduction in RSJD model: the continuous line
corresponds to a model in which the risk is not priced, the dotted
line to a variation of $+25\%$ of the volatility ($\sigma_1,
\sigma_2$ - left figure) and jump std ($b_1, b_2$ - right figure)
and the dashed line to a variation of $-25\%$. The parameters are
$\sigma_1=0.15, \sigma_2=0.05$, $b_1=0.1, b_2=0.2$, $r=0.05$,
$q_1=q_2=1$, $T=1$ and $S_0=1$. No substantial changes are
observed for this kind of variation of the other parameters
$\lambda_1=1, \lambda_2=0.1$, $a_1=0, a_2=-.1$, $q_i=1$.}
\label{RSJD_Red1Fig}
\end{center}
\end{figure}

Inside each model - GBM, JD, RSGBM, RSJD - we show the behavior of
the optimal hedging strategy obtained by changing the value of
some relevant parameters: this corresponds to specify different
levels of the market price of risk. In the figures
(\ref{GBM_Red1Fig}), (\ref{JDM_Red_mu1Fig}),
(\ref{JDM_Red_lam1Fig}, (\ref{RS_Red1Fig}) and
(\ref{RSJD_Red1Fig}) the optimal strike $K^*$ and the
corresponding value of the VaR are depicted together with the risk
reduction percentage
$$
R = 1 - \frac{\Var_{\alpha}(L^{h^*,K^*}_0)}{\Var_{\alpha}(L^u)}.
$$

It is apparent how the the strategy can be effective in reducing
risk. On the other hand, it can be noticed that a perturbation of
a single parameter results in a change of the optimal strategy.

\paragraph{(b) Misspecified Modeling.}

In order to explore the model sensitivity of the optimal hedging
strategy, we implemented the following exercise. We firstly fixed
a RSJD model by choosing a complete set of parameters. Then we
generated a set of call/put prices on which we calibrate the GBM
model: hence we run the optimal hedging strategy obtaining
$K_{GBM}^*, h_{GBM}^*$ and correspondingly the minimal VaR,
$\Var_{GBM}^*$. We finally calculated the probability
$$
\beta_{RSJD}=\P(L^{K^*,h^*} \geq \Var_{GBM}^* )
$$
under the RSJD model. Results are shown in Table
(\ref{Tab_optcompare1}) and (\ref{Tab_optcompare2}). Notice that
even when the optimal strategies are similar, the probability that
the portfolio loss exceeds the (optimal) VaR is greater than the
fixed level $\alpha=0.01$. Of course this behavior depends on the
choice of the parameters, but the underestimation produced by a
wrong model choice can be quite severe: see Table
(\ref{Tab_optcompare3}), where parameters from a real data set
were used (see \cite{Ram10}).

\begin{table}
\begin{center}
\begin{tabular}{|c|c|c|c|}
  \hline
       & Optimal strategy RSJD  & Optimal strategy GBM $(\hat \sigma)$ & $\beta_{RSJD}$  \\  \hline
T=.5   & 64.7442, 0.7197, 37.0189 & 65.6191, 0.8433, 35.3880 (0.2905)&  0.0132 \\
T=1    & 55.5928, 0.6165, 47.1767 & 57.1579, 0.7644, 44.4718 (0.2689)&  0.0148 \\
T=3    & 41.6851, 0.5294, 62.3356 & 43.5664, 0.7382, 58.6379 (0.2254)&  0.0157  \\
\hline
\end{tabular}
\end{center}
\caption{Optimal hedging strategy $(K^*,h^*,\Var^*)$ under the
simulated true model (first column), the fitted GBM model (second
column - estimated volatility) and the corresponding value of
$\beta_{RSJD}$. Here $\sigma_1=.3$, $\sigma_2=.05$, $\lambda_1=2$,
$\lambda_2=.8$, $a_1=0.0$, $a_2=0.0$, $b_1=0.08$, $b_2=0.15$,
$q_1=1$, $q_2=0.2$; furthermore $r=0.5\%$, $\alpha=0.01$ and the budget
constraint is $C=0.1$. \label{Tab_optcompare1}}
\end{table}

\begin{table}
\begin{center}
\begin{tabular}{|c|c|c|c|}
  \hline
       & Optimal strategy RSJD  & Optimal strategy GBM $(\hat \sigma)$ & $\beta_{RSJD}$  \\  \hline
T=.5   & 61.1841, 0.0581, 45.2341 & 63.3076, 0.0816, 41.3746 (0.3137) &  0.0166 \\
T=1    & 45.8347, 0.0833, 60.5069 & 52.7089, 0.0744, 52.6106 (0.3047) &  0.0255 \\
T=3    & 18.8056, 0.4015, 83.7630 & 32.7103, 0.0797, 72.5986 (0.2930) &  0.0640 \\
\hline
\end{tabular}
\end{center}
\caption{Optimal hedging strategy $(K^*,h^*,\Var^*)$ under the
simulated true model (first column), the fitted GBM model (second
column - estimated volatility) and the corresponding value of
$\beta_{RSJD}$. Here $\sigma_1=.3$, $\sigma_2=.05$, $\lambda_1=2$,
$\lambda_2=.8$, $a_1=0.05$, $a_2=-0.3$, $b_1=0.08$, $b_2=0.15$,
$q_1=1$, $q_2=0.2$; furthermore $r=0.5\%$, $\alpha=0.01$ and the budget
constraint is $C=0.01$. \label{Tab_optcompare2}}
\end{table}

\begin{table}
\begin{center}
\begin{tabular}{|c|c|c|c|}
  \hline
       & Optimal strategy RSJD    & Optimal strategy GBM $(\hat \sigma)$& $\beta_{RSJD}$  \\  \hline
T=.5   & 38.3721, 0.2497, 66.0564 & 60.0168, 0.0785, 44.8655 (0.3479)   & 0.1165 \\
T=1    & 26.6034, 0.4103, 76.3270 & 49.4859, 0.0737, 55.8926 (0.3320)   & 0.1304 \\
T=1.5  & 19.6884, 0.6506, 81.8069 & 41.9632, 0.0741, 63.4963 (0.3291)   & 0.1430 \\
\hline
\end{tabular}
\end{center}
\caption{Optimal hedging strategy $(K^*,h^*,\Var^*)$ under the
simulated true model (first column), the fitted GBM model (second
column - estimated volatility) and the corresponding value of
$\beta_{RSJD}$. Here $\sigma_1=.27$, $\sigma_2=.13$, $\lambda_1=6.8$,
$\lambda_2=.8$, $a_1=-0.13$, $a_2=-0.34$, $b_1=0.08$, $b_2=0.15$,
$q_1=6.5$, $q_2=0.002$; furthermore $r=0.5\%$, $\alpha=0.01$ and the
budget constraint is $C=0.01$. \label{Tab_optcompare3}}
\end{table}

\section{Conclusions}

In this paper we considered the problem of computing the quantiles
of a risky position described by a regime-switching jump-diffusion
dynamic model. The knowledge of generalized characteristic
function for this class of processes allowed us to use the Fourier
Transform methods to design an efficient algorithm for the
calculation of quantiles. With this same technique, we analyzed a
static hedging policy based on the constrained minimization of the
VaR of the option-hedged portfolio. Numerical examples showed the
impact of jumps and switching regimes on the optimal strategy in a
two-regime, gaussian jumps framework and moreover the risk of a
wrong model choice.

Some final comments can be briefly outlined. Firstly, notice that
different kind of jumps, as well as the number of regimes, can be
readily considered in our computational framework, such as the
double exponential Kou model (\cite{Kou02}). Furthermore, the
analysis of the hedging strategy is fairly general, that is it can
be applied to any dynamical model for which Fourier transform
methods are viable, for example it can be extended to
Variance-Gamma or Bates models. Finally, besides the choice of
different dynamic models, it would be interesting to consider
alternative risk measures, such as the Conditional Value at Risk
(CVaR). This is certainly less commonly used in finance industry,
but it is widely used in insurance industry being a coherent,
convex and stable risk measure (see \cite{ADEH}).

%\newpage
%%%%%%%%%%%%% BIBLIO %%%%%%%%%%%%%%%%%%%%%%%%%%%%%%%%%%%%%%%%%%

\newpage
%%%%%%%%%%%%%%%%% graphics %%%%%%%%%%%%%%%%%%%%%%%%%%%%%%%%%%%%%%%%%%%%%%%%%%%%%%%
%%%%%%%%%%%%%%%%%%%%%%%%%%%%%%%%%%%%%%%%%%%%%%%%%%%%%%%%%%%%%%%%%%%%%%%%%%%%%%%%%%

\begin{figure}
\begin{center}
\includegraphics[width=12cm,height=7cm]{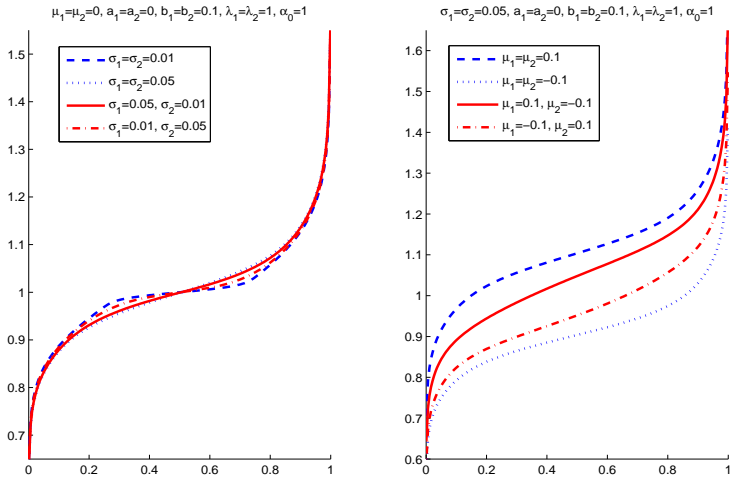}
\caption{\small RSJD model with gaussian jumps: switching
volatility (left picture) and drift (right picture). $T=1$.}
\label{RSJD_Pmu_sigmamix1Fig}
\end{center}
\end{figure}

\begin{figure}
\begin{center}
\includegraphics[width=12cm,height=7cm]{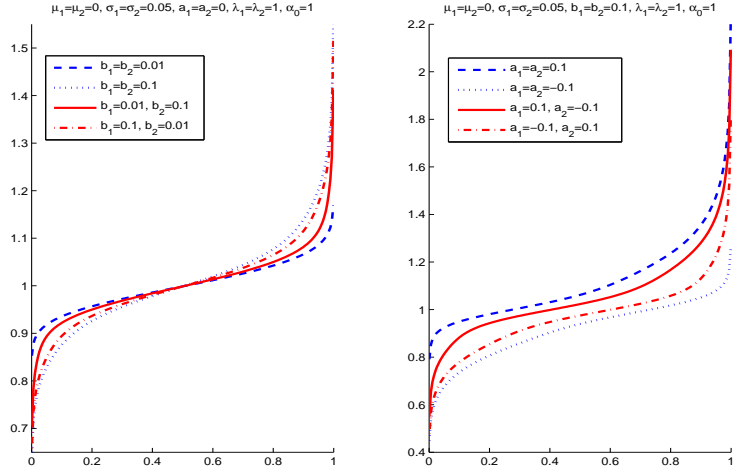}
\caption{\small RSJD model: switching variance (left picture) and
mean (right picture) of the gaussian jumps. $T=1$.}
\label{RSJD_Pa_bmix1Fig}
\end{center}
\end{figure}

\begin{figure}
\begin{center}
\includegraphics[width=12cm,height=7cm]{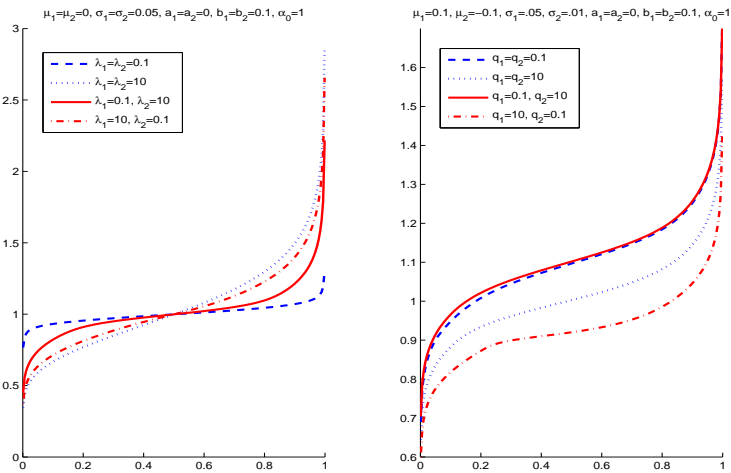}
\caption{\small RSJD model: switching jump arrival times
$\lambda_i$ (left picture) and intensity of the Markov chain $q_i$
(right picture). $T=1$.} \label{RSJD_Plam_qmix1Fig}
\end{center}
\end{figure}

%%%%%%%%%%%%%%%%%%%%%%%%%%%%%%%%%%%%%%%%%%%%%%%%%%%%%%%%%%%%%%%%%%%%%%%%%%%%%%%%%%
\end{document}